\documentclass{aa}

\usepackage[varg]{txfonts}
\usepackage{natbib,twoopt}

\newcommand{\rr}{\hbox{$r_{\rm Sloan}$}}
\newcommand{\bb}{\hbox{$B_{\rm Bessel}$}}
\newcommand{\uu}{\hbox{$U_{\rm spec}$}}
\newcommand{\bmr}{\hbox{$B_{\rm Bessel} - r_{\rm Sloan}$}}
\newcommand{\umb}{\hbox{$U_{\rm spec} - B_{\rm Bessel}$}}
\newcommand{\umr}{\hbox{$U_{\rm spec} - r_{\rm Sloan}$}}
\newcommand{\feh}{\hbox{[Fe/H]}}

\newcommand{\vv}{\hbox{$V$}}
\newcommand{\rh}{\hbox{$r_{\rm h}$}}
\newcommand{\rt}{\hbox{$r_{\rm t}$}}
\newcommand{\kms}{\hbox{km s$^{-1}$}}

\begin{document}
\title{A bag of tricks: Using proper motions of Galactic stars to identify
 the Hercules ultra-faint dwarf galaxy members
 \thanks{Based on data
 acquired using the Large Binocular Telescope (LBT) at Mt. Graham,
 Arizona.
 The LBT is an international collaboration among institutions in the
 United States, Italy, and Germany. The LBT Corporation partners are:  
 The University of Arizona on behalf of the Arizona university system;
 Istituto Nazionale di Astrofisica, Italy; LBT
 Beteiligungsgesellschaft, Germany, representing the Max-Planck
 Society, the Astrophysical Institute Potsdam, and Heidelberg
 University; The Ohio State University, and The Research Corporation,
 on behalf of The University of Notre Dame, University of Minnesota,
 and University of Virginia}
 }

\author{M.~Fabrizio\inst{1},
 G.~Raimondo\inst{1},
 E.~Brocato\inst{2},
 A.~Bellini\inst{3},
 M.~Libralato\inst{3,4,5,}\thanks{Visiting Ph.D. Student at STScI under the 2013 DDRF program.},
 V.~Testa\inst{2},
 M.~Cantiello\inst{1},
 I.~Musella\inst{6},
 G.~Clementini\inst{7},
 R.~Carini\inst{2},
 M.~Marconi\inst{6},
 G.~Piotto\inst{4},
 V.~Ripepi\inst{6},
 R.~Buonanno\inst{1,8}
 E.~Sani\inst{9},
 R.~Speziali\inst{2}}
\institute{INAF-Osservatorio Astronomico di Teramo, Via Mentore Maggini s.n.c., I-64100 Teramo, Italy\\
\email{fabrizio@oa-teramo.inaf.it}
\and INAF-Osservatorio Astronomico di Roma, Via Frascati 33, I-00040 Monte Porzio Catone (RM), Italy
\and Space Telescope Science Institute, 3700 San Martin Drive, Baltimore, MD 21218, USA
\and Dipartimento di Fisica e Astronomia, Universit\`a di Padova, Vicolo dell'Osservatorio 3, I-35122, Padova, Italy
\and INAF-Osservatorio Astronomico di Padova, Vicolo dell'Osservatorio 5, Padova, I-35122, Italy
\and INAF-Osservatorio Astronomico di Capodimonte, Salita Moiariello 16, I-80131, Napoli, Italy
\and INAF-Osservatorio Astronomico di Bologna, Via Ranzani 1, I-40127 Bologna, Italy
\and Dipartimento di Fisica, Universit\`a di Roma "Tor Vergata", Via della Ricerca Scientifica 1, I-00133 Roma, Italy
\and INAF-Osservatorio Astrofisico di Arcetri, Largo E. Fermi 5, I-50125 Firenze, Italy}

\date{Received .../ Accepted ...}

\abstract{Discovered in the last decade as overdensities of resolved
stars, the ultra-faint dwarfs (UFDs) are among the least luminous,
most dark-matter dominated, and most metal-poor galaxies known
today. They appear as sparse, loose objects with high mass-to-light
ratios. Hercules is the prototype of the UFD galaxies. To date, there
are still no firm constraints on its total luminosity due to the difficulty of
disentangling Hercules \emph{bona-fide} stars from the severe
Galactic field contamination.}
{To better constrain Hercules properties, we aim at removing
foreground and background contaminants in the galaxy field using the
proper motions of the Milky Way stars and the colour-colour diagram.}
{We have obtained images of Hercules in the \rr, \bb\ and \uu\ bands
with the Large Binocular Telescope (LBT) and LBC-BIN mode
capabilities. The \rr\ new dataset combined with data from the LBT
archive span a time baseline of about 5~yr, allowing us to measure 
proper motions of stars in the Hercules direction for
the first time. The \uu\ data along with existing LBT photometry allowed us 
to use colour-colour diagram to further remove the field contamination.}
{Thanks to a highly-accurate procedure to derive the \rr-filter
geometric distortion solution for the LBC-red, we were able to
measure stellar relative proper motions to a precision of better than
5~mas~yr$^{-1}$ down to \rr$\simeq$22~mag and disentangle a
significant fraction ($>$90\%) of Milky Way contaminants. We ended
up with a sample of 528 sources distributed over a large portion of the
galaxy body ($\sim$0.12~deg$^2$). Of these sources, 171 turned out
to be background galaxies and additional foreground stars from the
analysis of the \umb\ vs. \bmr\ colour-colour diagram. This leaves us
with a sample of 357 likely members of the
Hercules UFD. We compared the cleaned colour-magnitude
diagram (CMD) with evolutionary models and synthetic CMDs,
confirming the presence in Hercules of an old population
($t=$12$\pm$2~Gyr) with wide spread metallicity
($-3.3$$<$\feh$<$$-1.8$).}
{Our procedure to estimate star proper motions proved to be a very
effective way to identify likely members for stellar systems
as far as 130~kpc. The present selection, based on both proper
motions and colour-colour diagram, provides a robust identification of
Hercules members and a new target list for more spectroscopic
investigations.}


\authorrunning {M. Fabrizio et al.}
\titlerunning {Proper-motion of Galactic
 stars to identify \emph{bona-fide} members of the Hercules UFD }

\maketitle

\section{Introduction}

In the last decade, the ultra-faint dwarf (UFD) galaxies discovered
around the Milky Way (MW) and Andromeda spirals have provided
new insight in our knowledge to the principles that govern galaxy
formation processes. The UFDs show a number of remarkable
differences with respect to the "classical" dwarf spheroidal galaxies
(dSphs). They are generally fainter than the previously known
spheroidals with surface brightnesses typically less than
27~mag~arcsec$^{-2}$; thus, they are given 
the name "ultra-faint dwarfs". Although
UFDs have dimensions comparable to those of the classical dSphs
(the typical half-light radius is \rh$\geq$100~pc), their absolute \vv\
magnitude ($M_V$) is similar to that of the bulk of Galactic globular
clusters (GGCs, $M_V$$\approx$$-7$~mag), and some of them are
even less luminous with absolute magnitudes as low as
$M_V$$\approx$$-2$~mag \citep[see e.g.][]{clementini12}. All MW
UFDs host an ancient population as old as about 10~Gyr and, with
only very few exceptions  \citep[e.g.][]{clementini12,okamoto12}, their
CMDs resemble those of metal-poor Galactic clusters
\citep{belokurov07,brown12} with no evidence for
intermediate-age stellar generations, which are often observed
among the classical dSphs instead \citep[e.g.][]{monelli03,bellazzini06}.
Measurements of the UFDs internal velocity dispersion reveal
surprisingly high values ($\sim 3\div10$~\kms) in comparison to the
luminosity \citep[e.g.][]{simon07,walker09}. Therefore, in contrast to 
GGCs and even more so in classical dSphs, UFDs appear to be
dark-matter dominated with mass-to-light ratios that can reach as large
as $M/L$$\sim$$10^{2\div3}$~M$_\sun$/L$_\sun$.

\begin{figure*} 
\center
\includegraphics[trim=2.8cm 2cm 4.5cm 11.3cm,width=1\columnwidth,clip=true]{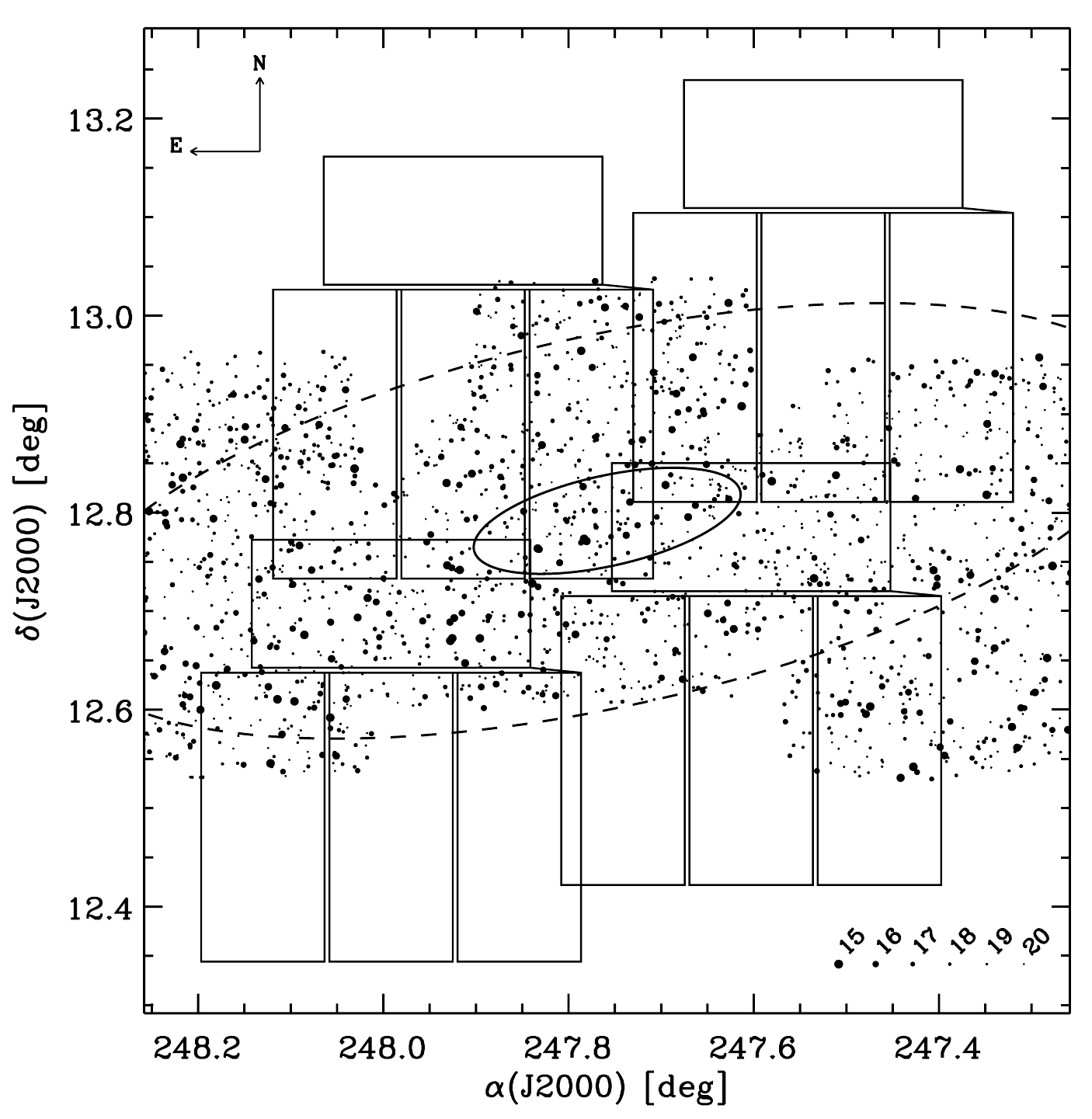}
\includegraphics[trim=3.5cm 2cm 3.8cm 11.3cm,width=1\columnwidth,clip=true]{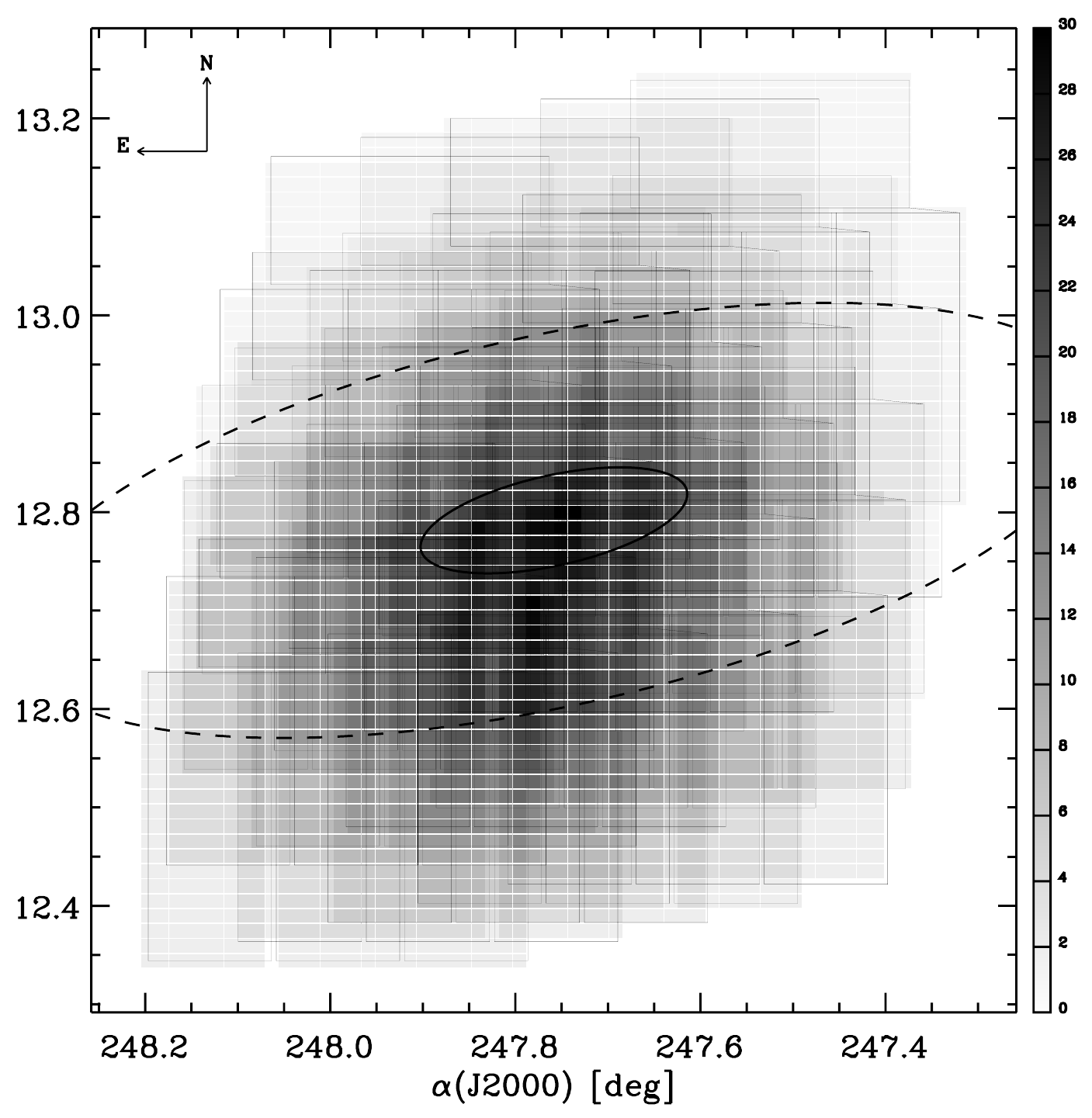}
\caption{Left: Map of Hercules UFD showing the regions covered by the
proprietary and archive LBT observations used in the present study. The
solid lines display the contours of the most external LBC pointings of the
dither pattern adopted during the April 2013 observations.
The black points show the position of the photometric sources in the
central and lateral fields observed by \citet{sand09}. Symbol size is
inversely proportional to the source \rr\ magnitude. The solid and dashed
ellipses mark the galaxy half-light and tidal radii, respectively.
Right: Footprints of the 5$\times$5 dither pattern adopted during the
April 2013 observations. The depth-of-coverage map for the \rr-images is
also shown in grey scale.
\label{fig:fields}}
\end{figure*}

Dwarf spheroidal galaxies have long been thought to be candidate
building-blocks in the hierarchical scenario of galaxy formation
\citep[e.g.][]{grebel05}, where galaxies form via cooling and
condensation of gas in dark-matter halos \citep{white78}. However, the
role played by the classical dSphs in shaping the halos of their
massive parent galaxies still remains quite controversial, a major issue
being the small number of classical dSphs observed in the
Local Group (LG). On the other hand, the number of UFDs has
increased constantly in the last decade and completeness estimates
suggest that many more of these faint satellites are still to be
discovered in the LG \citep{tollerud08}. They may eventually fill the
gap between observed number of satellites and number of relics
predicted by the $\Lambda$CDM models, the so-called "missing
satellite problem" \citep{moore99,klypin99}.

The Hercules UFD was discovered by \citet{belokurov07} from the
analysis of Sloan Digital Sky Survey (SDSS) data. Literature estimates
for its distance range from 132 kpc \citep[e.g.][]{coleman07} to
147~kpc \citep[e.g.][]{aden09spec} with a value of 132$\pm$6~kpc
measured from the galaxy RR Lyrae stars \citep[][]{musella12}. It has
an absolute magnitude of $M_V$$=-6.6$$\pm$0.3~mag, which
places the galaxy among the brightest MW UFDs, and a surface
brightness of $\mu_V$=27.2$\pm$0.6~mag~arcsec$^{-2}$
\citep{martin08}. Since the earliest ground-based observations, it was
recognized that the galaxy CMD resembles that of an old and
metal-poor GGC \citep[$t$$\gtrsim$10$\pm$2~Gyr and
Z$\sim$10$^{-4}$, see e.g.][] {sand09,musella12}, as recently confirmed
by \citet{brown12} with \emph{Hubble Space Telescope} (\textit{HST})
data.
\cite{martin08} measured the galaxy half-light radius (\rh=8.6\arcmin),
and \cite{sand09} estimated the tidal radius (\rt=35.25\arcmin).
Hercules appears highly elongated with an ellipticity $e$$\gtrsim$0.65
\citep{coleman07,martin08} and shows indications of a possible tidal
disruption \citep[e.g.][]{martin10}. \citet{deason12} suggested that
some hot horizontal branch (HB) stars once belonged to Hercules may
have been tidally stripped and are now unbound. Thus, Hercules may
not be a simple bound system in equilibrium
\citep{aden09spec,martin10,deason12}. If so, its total mass cannot be
gauged via measurement of the overall velocity dispersion and size
\citep[see e.g.][]{klimentowski07,lokas09},
as the streaming of unbound stars could, in principle, inflate the
observed mass-to-light ratio.

Unfortunately, the line of sight towards Hercules is heavily
contaminated by external sources, making it hard to establish
membership from the CMD alone \citep[e.g.][]{musella12}. 
Hercules, as all UFDs, is a very loose stellar system
($\rho$=1.7$\times$10$^{-2}$~M$_\sun$~pc$^{-3}$,
\citealt{walker09}); hence, background galaxies constitute a severe
source of contamination.
In addition, this UFD also suffers from a large contamination by MW
foreground stars. Even when radial velocities (RV) are
considered, the selection of \emph{bona-fide} members is made rather
uncertain by the mean velocity of the galaxy ($\sim$45$~\kms$,
\citealt{aden09spec}) being embedded into the velocity distribution of
foreground MW dwarf stars, coinciding with the velocity of the thick
disk.
For this reason, an estimate of Hercules mass based only on
kinematic methods is highly uncertain, as shown by
\citet{aden09mass}. Adding a selection based on the Str\"{o}mgren
$c_1$ index, \citeauthor{aden09mass} reduced the radial-velocity
dispersion of Hercules stars and decreased the galaxy mass within
the half-light radius by a factor of $\sim$1/4. The rescaled mass is
much lower than the \emph{common mass scale} for MW satellites
\citep{strigari08}, suggesting that Hercules does not share the halo
properties seen in the classical dSphs and other UFDs. The Galactic
contamination in the direction of Hercules is a severe problem that
needs to be settled before attempting to derive metallicity and radial
velocity distributions of Hercules's stars and, in turn, constrain the
galaxy mass and luminosity.

Due to the importance of weeding out any external contaminants, an
accurate method to determine the star membership is mandatory to
study the stellar population content and the structure of such a
poorly-populated and heavily-contaminated system.
In this paper, we study the star membership through the analysis of the
proper motion (PM) of stars in the Hercules field. Our aim is to remove
the foreground contamination by separating Galactic from Hercules
stars, according to their PMs. Up to now, such measurements have
been conducted in a handful of dSphs using \textit{HST} data
\citep[e.g.][]{piatek03,piatek07,massari13}, while the Fornax galaxy is
the only target studied with ground-based data \citep{mendez10}.
However, previous works dealt with limited areas of those
galaxies, due to the small field-of-view (FoV) of the detectors.
Here, we use
the Large Binocular Telescope (LBT), whose FoV covers well beyond
the half-light radius of Hercules (\rh=8.6\arcmin, \citealt{martin08}). We
combine new observations specifically obtained for this purpose with
archival images and found our membership criteria on PMs and on the
colour-colour diagram that involve \bb, \rr, and \uu\ data.

The paper is organized as follow: \S~\ref{sec:obsana} describes our
procedures for the data analysis and to measure PMs and
photometry. The results are discussed and compared to previous
membership analysis in \S~\ref{sec:membership}, while implications
for the galaxy stellar population content and for the metallicity
and radial-velocity distributions are presented in
\S~\ref{sec:stellarpopulation}. A summary and
conclusions section close the paper (\S~\ref{sec:conclusions}).

\section{Observation and data reduction}
\label{sec:obsana}

\subsection{Observations}
\label{sec:observations}

The Hercules UFD galaxy
($\alpha$=16$^\mathrm h$31$^\mathrm m$02$^\mathrm s$.0,
$\delta$=12$^\circ$47\arcmin25\arcsec.6, J2000.0)
was observed with the Large Binocular Camera
\citep[LBC,][]{giallongo08} at the LBT located on Mount Graham in
Arizona. The LBC is a wide field imager (4 CCDs, 2K$\times$4.5K
pixels each) with a FoV of $\sim$23\arcmin$\times$23\arcmin\ and a
resolution of 0.225\arcsec~pix$^{-1}$. We employed both archival and
proprietary data to have the longest available time baseline. A
log of the observations used in this work is reported in
Table~\ref{tab:obslog}.

The 2007 dataset covers only the central part of the galaxy and was
collected during the LBT science demonstration time (March 2007,
\citealt{coleman07}) with the LBC-blue, the only active "eye" at that
time. Images were taken through \bb- and \rr-filters with almost no
dither. An accurate geometric-distortion (GD) solution is available for
the \bb\ filter, but not for the \rr\ filter. It was not possible to
self-calibrate the GD of the \rr\ filter due to the lack of an adequate dither
strategy. For this reason, we selected and used only
images in the \bb\ filter from this dataset.

The archival \rr\ exposures are part of the dataset collected
between May 29th and June 1st 2008 with the LBC-red and published
in \citet{sand09}, from which we selected the central and lateral fields
(see Fig.~\ref{fig:fields}).

The archival images were combined with 25 new images
acquired on April 4th 2013\footnote{INAF-DT n.38, P.I.:\ E.~Brocato}.
We adopted a large dither pattern (70\arcsec, $\sim$30\% of the FoV)
specifically designed to calibrate the GD in the \rr-filter of the LBC-red.
The right panel of Fig.~\ref{fig:fields} shows the dither pattern and the
depth-of-coverage map of the proprietary \rr-exposures. The total area
covered is 0.69~deg$^2$. Images were collected using both the
LBC-blue and LBC-red arms in \rr\ (39 exposures), \bb\ (31), and \uu\ (8).

The new observations completely cover the central field and partially
overlap the lateral fields of \citet{sand09}, as shown in the left panel of
Fig.~\ref{fig:fields}.

\begin{table} 
\center
\tiny
\caption{Log of LBT Observations used in the present study.}
\label{tab:obslog}
\begin{tabular}{lcccccc}
\hline
\hline
UT Date & LBC & Filter & Exp.time & Seeing & Dither & Field \\
&&&(s) & (arcsec) && \\
\hline
2007 Mar 17 & Blue & \bb & 6$\times$300 & 0.9 & Small & Cent \\
\\
2008 May 29 & Red & \rr & 6$\times$300 & 0.9 & Small & East\\
2008 May 30 & Red & \rr & 6$\times$300 & 0.9 & Small & West\\
2008 May 31 & Red & \rr & 6$\times$300 & 1.2 & Small & Cent\\
2008 Jun 01 & Red & \rr & 5$\times$300 & 0.9 & Small & Cent\\
\\
2013 Apr 04 & Red & \rr & 39$\times$60 & 0.9 & Large & Cent\\
2013 Apr 04 & Blue & \bb & 31$\times$60 & 1.1 & Large & Cent\\
2013 Apr 04 & Blue & \uu & 8$\times$60 & 1.2 & Large & Cent \\
\hline
\end{tabular}
\end{table}

\subsection{Data reduction}
\label{sec:datareduct}

Data were reduced using the procedures described in
\citet{bellini10m67}. A main asset of our reduction process was the
derivation of an ad-hoc GD correction for the LBC-red in the \rr-filter,
starting from the recipes described in \citet{bellini10gd} and improving
the polynomial solution that follows prescriptions provided in
\citet{libralato14}. Here, we briefly outline the key points of the
reduction process while referring to the aforementioned papers for
more details.

\subsubsection{Astrometric and photometric reduction}
\label{sec:reductions}

The pre-reduction procedure included standard operations such as
de-biasing, flat-fielding, pixel-area correction, and cosmic rays-removal
\citep{bellini10m67}. The reduction software is able to find bright,
isolated star-like sources in each CCD of each exposure and obtain
spatially varying empirical point-spread functions (PSFs in an array of
3$\times$4 elements for each chip).
These PSFs were then used to measure positions and fluxes of
sources in each chip of each exposure that were then collected into
individual catalogues.

\begin{figure*} 
 \centering
 \includegraphics[width=1.5\columnwidth]{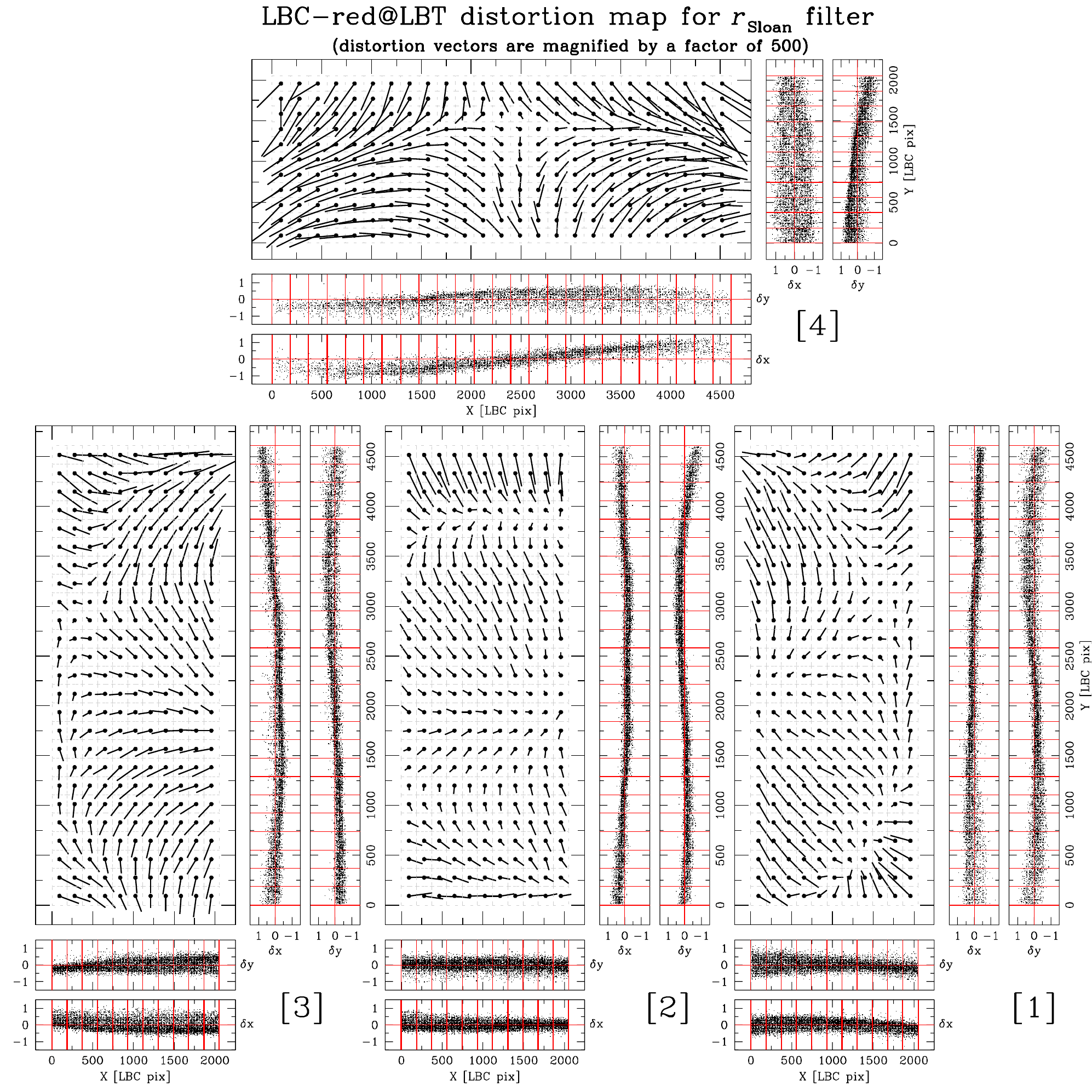}
\caption{Residual trends of the four chips for
 uncorrected star positions and single residual trends along the X
 and Y axes. The size of the residual vectors is magnified by a
 factor of 500. Units are expressed in LBC-red raw
 pixels.}
 \label{fig:maps1}
\end{figure*}

\subsubsection{The master frame}

All stellar positions in the \bb-filter catalogues were corrected
using the \bb-filter geometric-distortion correction available from
\citet{bellini10gd}. We cross-identified stars in common
between catalogues and derived average position and magnitude
for each source onto a distortion-free reference system (hereafter, the
master frame). To build the master frame, we proceeded as follows.
We started from the GD-corrected positions of the first catalogue and
identified all catalogues that have stars in common. We then
transformed the stellar positions of these overlapping catalogues to
the first-catalogue reference system by means of conformal
transformations (two rigid shifts in the two coordinates, one rotation
and one change of scale), using only bright, unsaturated stars in
common to compute transformation coefficients. The average
transformed positions and fluxes of common stars defined a new,
larger, and improved master frame. We iteratively increased and
improved the master frame by adding additional overlapping
catalogues.

Once stars from all catalogues have been added to the master frame
list, we improved single-catalogue transformations by means of more
general, 6-parameter linear transformations (to minimize the residuals
in the linear terms). The final master frame contains average
positions and fluxes of all stars measured in at least three
exposures.

\subsubsection{\rr-filter geometric distortion solution for the LBC-red}
\label{sec:gd}

\begin{figure*} 
 \centering
 \includegraphics[width=1.5\columnwidth]{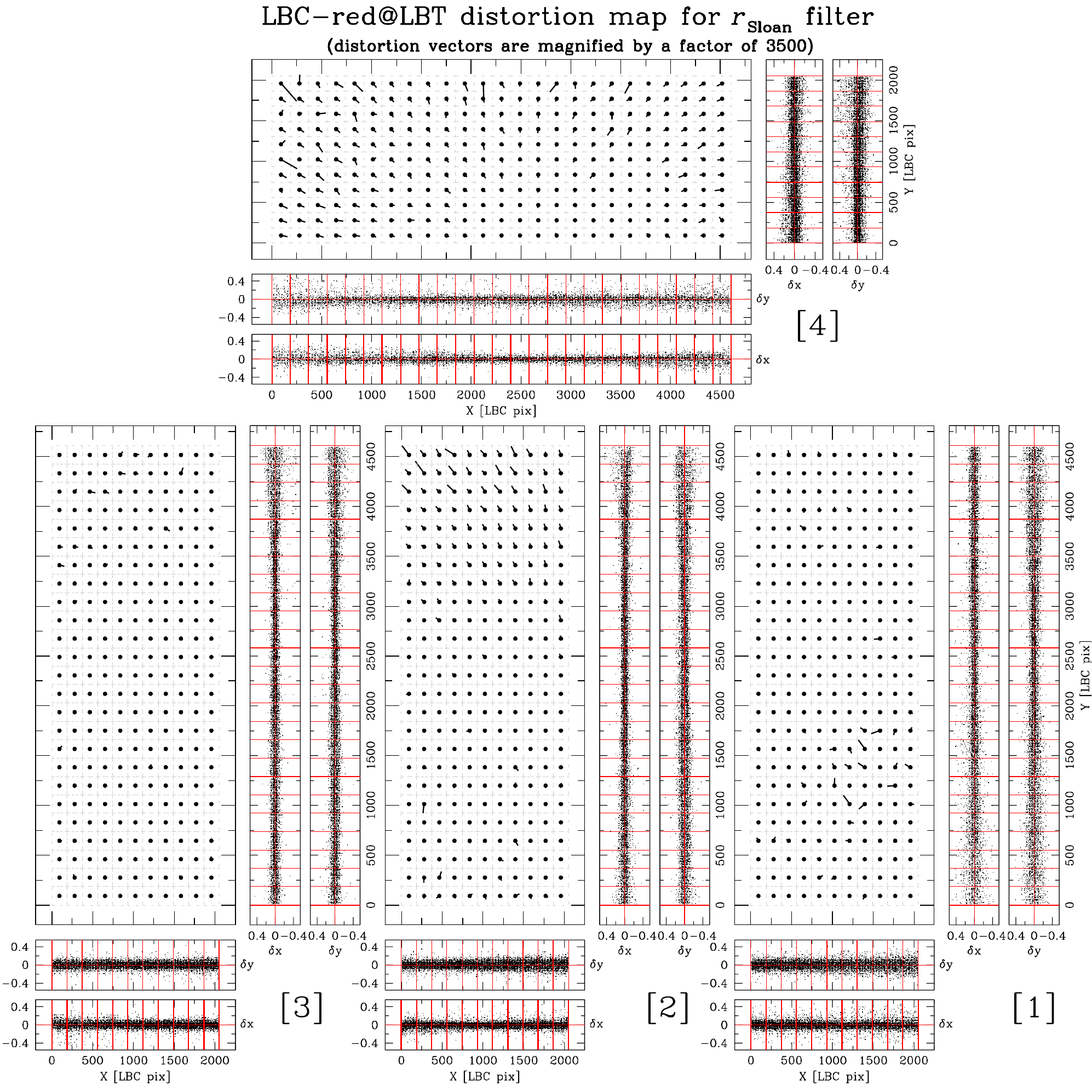}
 \caption{Positional residuals after all corrections have been
 applied. The size of the vectors is now magnified by a
 factor of 3500.}
 \label{fig:maps2}
\end{figure*}

The blue and red LBC arms have the same optical design, and their
cameras are made by the same detectors arranged in the same way.
This means that both the blue and the red cameras
have a similar geometric distortion to the 0-th order, and, therefore, 
the distortion solution found for the blue-arm should represent a good first-guess solution also for the red-arm.

Hence, to derive the geometric-distortion solution of the \rr\ exposures
for the LBC-red, we started by correcting the stellar positions of the
2013 dataset with the \vv-filter GD correction of \citet{bellini10gd}.

We cross-identified stars in different catalogues to put them in the
same reference frame of the \bb-filter master frame by means of
conformal transformations. Even if the \vv-filter correction for the
LBC-blue is not perfectly adequate to correct the \rr-filter LBC-red
images, systematic errors on their positions have a random amplitude
from one exposure to the other due to the adopted observing strategy
(dither pattern). Hence, the average \rr\ star positions in the master
frame provide a good first-guess approximation of their true positions in
a distortion-free frame.

The GD solution for the \rr-filter of the LBC-red is made by three parts:
$(i)$ a linear transformation to put all images on the master frame;
$(ii)$ a polynomial correction; and $(iii)$ a table of residuals.

The polynomial correction was mostly performed, as described in
\citet{bellini10gd}. The only difference is that we used a fifth-order
polynomial instead of a third-order one to fit the residual trends.
At the first iteration, the cubic part of the
polynomial was the same of the \vv-filter third-order polynomial of
\citeauthor{bellini10gd}, while higher-order terms were set to zero.
Then, we iterated the procedure to derive new polynomial coefficients
and improve the master frame until the star positional r.m.s. did not
change significantly from one iteration to the next.

The last part of our correction consisted of four look-up tables (one
for each CCD) to minimize the residuals of the geometric distortion
left by the polynomial correction. We followed the method described in
\citet{bellini11} for the WFC3/UVIS detector of the \textit{HST}.

In our correction, we subdivided each chip of the 4-CCDs LBT mosaic
in 11$\times$25 elements. The table correction was built at
a given location of the chip using a bi-linear interpolation among the
surrounding grid points.
We iterated this procedure by applying 75\%\ of the suggested 
correction, building a new master frame, and computing new, improved 
grid points. The convergence was reached when the change in the 
suggested correction was negligible from one iteration to 
the next.

\begin{figure*}[t] 
 \centering
 \includegraphics[width=1.5\columnwidth]{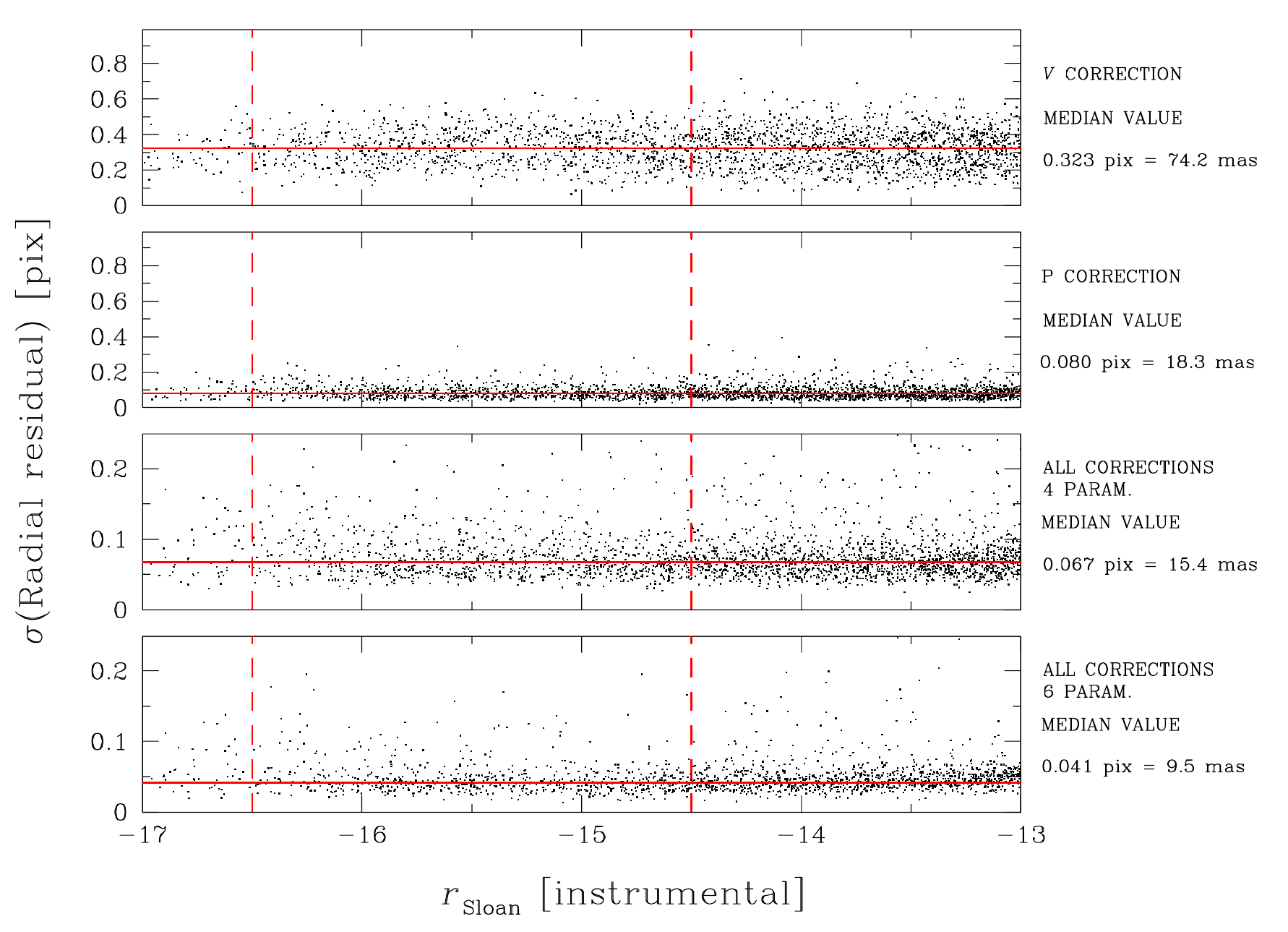}
 \caption{$\sigma$(Radial residuals) versus instrumental \rr\
 magnitudes after each step of our solution. The red solid
 horizontal lines show the median value of $\sigma$(Radial residual)
 between red dashed vertical lines.}
 \label{fig:1dm}
\end{figure*}

In Figs.~\ref{fig:maps1} and \ref{fig:maps2}, we show the distortion
maps before and after applying the distortion corrections, respectively.
Figure~\ref{fig:1dm} shows the
$\sigma$(Positional residuals), computed as in \citet{libralato14}
after each step of our correction. In the bottom panel,
we show the $\sigma$(Positional residuals) for
a master frame derived using general 6-parameter
linear transformations.
By applying all corrections, the $\sigma$(Positional residuals) improve
from $\sim$74.2~mas to $\sim$9.5~mas.

\subsubsection{Geometric-distortion corrected and cleaned catalogues}
\label{sec:finalcatalogue}

\begin{figure*} 
\center
\includegraphics[trim= 0.5cm 0.5cm 1cm 7cm, clip=true,width=1.8\columnwidth]{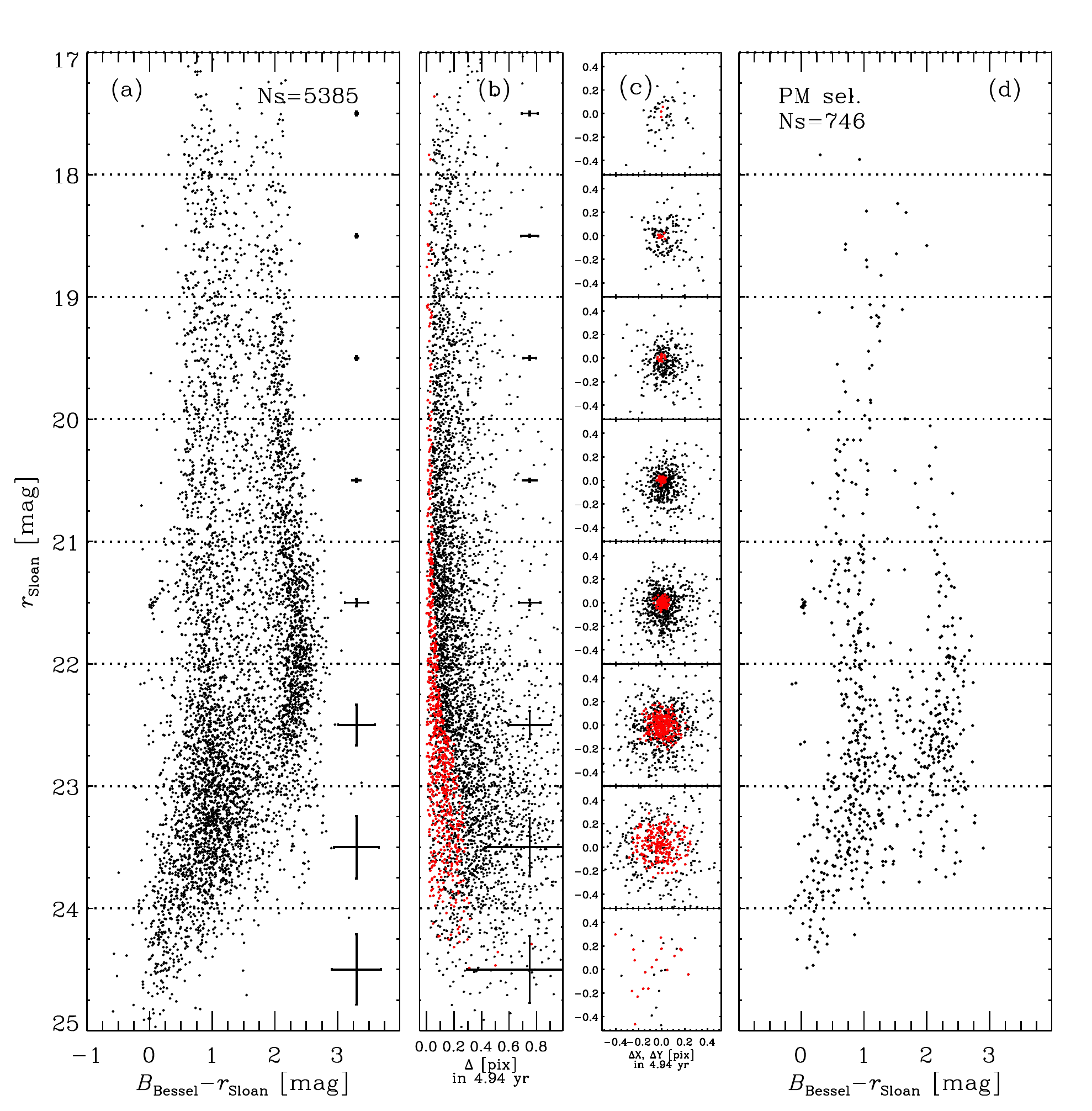}
\caption{Panel (a): \rr\ vs. \bmr\ CMD-cleaned, according to the
photometric selection criteria described in the text. Mean errors are
also shown on the right for each magnitude bin. Panel (b):
Displacement based on the \rr-filter as a function of magnitude. Mean
errors are also shown. Red symbols mark proper-motion (PM)
selected stars (see text). Panels (c): Vector-point diagrams for each
magnitude bin. Symbols are the same as in panel (b).
Panel (d): \rr\ vs. \bmr\ PM-selected CMD.
\label{fig:pm_cmd}}
\end{figure*}

To build the final master frames for the \uu\footnote{The GD correction
in \bb\ was adopted also for the \uu filter.}, \bb, and
\rr\ filters, we cross-identified all stars in our GD-corrected
catalogues using 6-parameter linear transformations to absorb
uncorrected distortion and atmospheric effects
\citep[see e.g.][]{libralato14}
and to further improve the general astrometric accuracy.

The master lists (one for each filter) were purged from false
detections as in \citet{anderson08a}. We selected stellar sources
according to their photometric errors and the quality of the PSF-fit
(\textsf{QFIT}, derived from the absolute value of the residuals of
the PSF-fit for each star scaled by the flux) by considering the 65th
percentile of the quoted quantities as a function of magnitude. We
ended up with master lists of 15217 sources in \rr, 7700 in \bb\ and
5759 in \uu.

\subsubsection{Astrometric and photometric calibration}
\label{sec:calibration}

The astrometric calibration was also derived with the purpose
of obtaining the match with the SDSS \citep[DR9,][]{ahn12}
and the Sand et al.'s catalogues.
We used the UCAC~4 catalogue \citep{zacharias13} to convert 
our pixel-based coordinates into equatorial coordinates 
by cross-identifying the stars
in common between the two catalogs using 6-parameter linear
transformations.

The photometric calibration was performed using a set of
sources extracted from the SDSS catalogue.
The \bb\ magnitudes were obtained by applying the relations by
\citet{jordi06} and \citet{jester05}, as described in \citet{sand09}.
The \uu\ filter has a response curve very similar to the SDSS $u'$;
therefore, we calibrated it against the SDSS
magnitudes without any further colour transformation.
The zero point and the colour term of the calibration relations were
obtained using only the best stellar-type sources
(i.e. \textsf{QFIT}$<$0.15) where the r.m.s. of the
photometric residuals around the
mean magnitude is smaller than 0.03~mag.
For the \rr\ and \uu\ filters, a zero point was sufficient to calibrate the
instrumental magnitudes, and no significant colour term was found. For
the \bb\ filter, both \umb\ and \bmr\ colours were used to derive
calibration relations. We obtained a good fit with both colours but 
used the equation in \bmr\ in the end , which has
smaller scatter. The calibration results were checked against
\citet{sand09} for the two filters in common (\bb, \rr), finding a
close agreement with an r.m.s. scatter below $\sim$0.03~mag
for magnitudes brighter than \bb$\sim$20~mag
and \rr$\sim$21~mag. 
For the \uu-filter, the check was performed by
comparing the calibration results obtained from the \umb\ and 
the \umr\ colours to fit the colour
term, which was negligible in both cases, as already mentioned. The
check was performed by using a set of $\sim$1500 stars from the SDSS
survey having a magnitude brighter than $u'$$\sim$20~mag in the colour
range 1$<$$u'-r'$[mag]$<$3. The results of the fit provide calibrated
\uu\ magnitudes that are in good agreement with each other. In
the following, we adopt the calibrated \uu\ magnitudes coming from
the \umr\ transformation equation.

\subsection{Proper motions}
\label{sec:pm-analysis}

The \rr\ images were used to measure the stellar proper motions
(PMs). The Hercules \rr-band datasets define a time baseline of
4.94~years and cover a common area of 0.30~deg$^2$. To measure
the stellar displacements between the two epochs, we followed the
prescriptions given in \citet{anderson06}. 
To do this, it is mandatory to select a set of reference objects 
against which to compute relative PMs. An absolute reference frame is 
not required here. We measure star motions with respect to other 
stars in our field.
From basic considerations on the expected internal motions of stars 
belonging to Hercules as opposed to the Galactic field stars, we can 
expect that the internal velocity dispersion of the formers is much 
smaller than that of the latters. Indeed, according to the internal velocity 
dispersion measured by \citet{aden09spec} and the galaxy distance, we 
expect the Hercules stars to have an internal velocity dispersion of $\sim
$0.01~mas~yr$^{-1}$. This is a factor of $10^3$ smaller than our
random measurement errors ($\sim$10~mas~yr$^{-1}$). Furthermore,
according to what is observed for other dSphs located at similar
distance, such as Carina \citep[$\sim$100~kpc,][]{piatek03} and
Fornax \citep[$\sim$135~kpc,][]{piatek07}, we expect an absolute PM
of the Hercules stars of the order of tenths of mas~yr$^{-1}$.
Therefore, Hercules stars represent the best choice to define our
reference list.

We started by identifying Hercules stars within an ellipse with a
semi-major axis of 7\arcmin\ centered on the galaxy center,
ellipticity, and inclination values from \citet{martin08} (to mitigate
the fraction of field contaminants). Stars within this region were
then plotted on the \rr\ vs.\ \bmr\ CMD, on which we isolated two
regions containing \emph{bona-fide} Hercules red giant branch
(RGB) and HB stars.
The CMD selections were then extended to the whole FoV.
The initial list of reference stars includes 323 objects.

We used this starting reference list to transform stellar positions of
each catalog onto the master frame following the local transformation
approach described in \citet{anderson06}. 
This allows us to map the coordinate system of one frame into 
that of another and transform the position of each individual star 
measured in one frame into that of the other.
Because of the low stellar density in the Hercules field, we used 
the nearest 25 reference stars found on the same chip as the target star 
in both images. Of these 25 stars, we reject the five with the largest 
transformation residuals. The final best 20 reference stars will then 
define the local transformation parameters that map the coordinate 
system of one frame into that of another.
Then, for each star, we computed the displacement on the master
frame as the average of the displacements obtained by comparing
each pair of transformed positions (one from the first epoch and one
from the second epoch).
Once we have obtained a first estimate of the stellar PMs, we can
improve the reference list by rejecting those stars, which motion is
not consistent with the Hercules mean motion (i.e. with a displacement
larger than 1~pixel). We iterated the process of improving the
reference list and measuring more accurate PMs three times. A further
iteration proved to offer negligible improvements. The final list of
reference stars includes 255 objects.

Because of the low surface density of Hercules stars and the 
severe Galactic field star contamination, some of the reference stars 
used in the local transformation between epochs are actually not 
members of Hercules. In particular, the fraction of Galactic stars is 
expected on the basis of Besan\c{c}on model to increase from $\sim
$20\%\ to $\sim$50\%, which moves from the center to the outer regions 
of the field. However, the inclusion of these stars in the reference list does 
not sizably affect the astrometric reference, since the mean PM of such 
stars is comparable to that of Hercules within the measurement errors 
but with a larger dispersion.

We were able to measure PMs for 5385 objects that have both \rr\ and
\bb\ photometry. Panel (a) of Fig.~\ref{fig:pm_cmd} shows the
\rr\ vs. \bmr\ CMD of these sources.
Similar, but less accurate, results were obtained using probable
background galaxies with 23.5$<$\rr\ [mag]$<$24.5 and
$-0.5$$<$\bmr\ [mag]$<$1.0 as the
reference list. The disadvantage of this approach is in the much
lower signal-to-noise of these sources (higher random errors) that
leads unavoidably to a poorer accuracy in the PM determination.

\subsubsection{Proper motion selection}
\label{sec:pm_sel}

The displacements ($\Delta=\sqrt{\Delta {\rm X}^2+\Delta {\rm Y}^2}$)
for the 5385 stars, which have a PM measure, are reported as black
dots in panel (b) of Fig.~\ref{fig:pm_cmd}. The mean uncertainties of
magnitude and displacement in each magnitude bin are also shown.
Unlike what is commonly found for Galactic stellar clusters,
panel (b) does not show a clear dichotomy in the PM star distribution.
Therefore, to distinguish field stars from Hercules members,
we can only select those stars with a PM compatible with a null value.
For this reason, we took the dependence of the PM
uncertainty on magnitude into account by computing a running 65th 
percentile of the r.m.s. displacement ($\sigma_\Delta=
\sqrt{\sigma_{\Delta {\rm X}}^2+\sigma_{\Delta {\rm Y}}^2}$) as a
function of the \rr\ magnitude. We, then, adopted its half as PM
selection radius ($R_{\rm sel}$). Moreover, we chose only stars that
have a number of couples of measurements ($N_{\rm c}$) larger than
50. These selection criteria represent a compromise between
including contaminants that have velocity equal to the Hercules' mean
PM and missing galaxy members with larger PM errors. The resulting
selections are shown as red symbols in panel (b). The same colour
coding is adopted in panel (c) of Fig.~\ref{fig:pm_cmd}, where we
show the vector-point diagram (VPD) for each magnitude bin. In other
words, the selected stars have a displacement lower than the
(magnitude dependent) uncertainty; thus, their displacement is
assumed to be nil.

The final PM-selected catalogue consists of 746 sources,
and we show their CMD in panel (d) of Fig.~\ref{fig:pm_cmd}.
The RGB and HB of Hercules stars are clearly
defined, compared to the unselected CMD in panel (a). In both panels,
the main sequence turn-off of Hercules (\rr$\sim$24~mag,
\citealt{sand09}) is not reached with an adequate signal-to-noise ratio,
although we have used the same data of \citet{sand09}. This effect is
caused by the lower (by a factor of 5) total exposure time of the
second-epoch data compared to the first-epoch, and it is even
enhanced by the different approach to the data reduction and
analysis. \citeauthor{sand09} performed their photometry on the
co-added images; we instead measured the stars positions and fluxes
on the individual exposures.

\begin{figure} 
\center
\includegraphics[trim= .2cm .2cm .5cm 7.5cm, clip=true,width=.99\columnwidth]{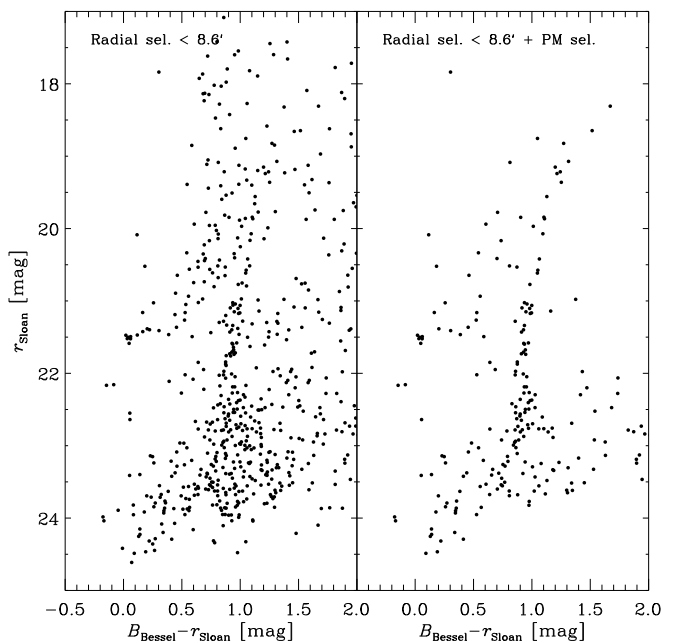}
\caption{CMD of stars within an elliptical region with \rh=8.6\arcmin\
before (left) and after (right) applying the PM selection procedure.
\label{fig:radial_comp}}
\end{figure}

The PM selection allows us to disentangle a significant fraction of MW
stars. To better highlight our results in Fig.~\ref{fig:radial_comp}, we
compare the CMD within Hercules half-light radius before and after
the PM-selection procedure. The left panel of Fig.~\ref{fig:radial_comp}
shows that a selection merely based on the stars
radial distribution still suffers from a large contamination by field stars.
This is significantly reduced by PM-motion selecting Hercules
members (right panel of Fig.~\ref{fig:radial_comp}). Nevertheless,
panel (d) of Fig.~\ref{fig:pm_cmd} shows that a Galactic sequence at
colour 1.9$<$\bmr\ [mag]$<$2.8 is still present in our PM-selected
CMD. To provide a rough estimate of the residual Galactic
contamination, we selected two boxes on this sequence and compared
the star counts with the unselected catalogue. 
In the first box, which are located
around the HB magnitude level (21$<$\rr\ [mag]$<$22), we found 
that about 8\%\ of Galactic stars survive our proper motion selection, 
while 25\%\ are in the second box, located at a
fainter magnitude level (22$<$\rr\ [mag]$<$23).
We note that similar fractions are expected
from a Galactic simulation taken from the Besan\c{c}on stellar
population synthesis models \citep{robin03,robin04} after applying the
same PM-selection criteria described above.
The same simulation allows us to estimate the residual fraction
of Galactic stars expected
at the same levels of magnitude as above, but for the Galactic blue plume
in the colour range $-0.5$$<$\bmr\ [mag]$<$1.8. They are
expected to be $\sim$10\%\ and 40\%, respectively.
In Table~\ref{tab:hercstars}, we list the final catalogue of 528
PM-selected sources. From left to right, we report
the source identity (Col.~1), right ascension, and declination (Col.~2--3),
\uu, \bb\ and \rr\ magnitudes with their uncertainty (Col.~4--6),
displacement with its uncertainty (Col.~7).

We recall that the selection described above is based on the assumed
selection radius for the allowed displacement and on the number of
couples. We performed several tests to evaluate the
dependence of our results on these quantities. We found that a
30\%\ increment (decrement) of $R_{\rm sel}$ produces an increase
of 25\%\ (decrease of 40\%) in the number of sources, while the
contamination increases (decreases) more than 35\%.
On the other hand, the number of sources increases by 70\%\ for
$N_{\rm c}$=10, but the residual contamination doubles. In
contrast, by adopting $N_{\rm c}$=100, the field contamination and
sources are reduced to half the number. We can conclude that the
parameters adopted in Fig.~\ref{fig:pm_cmd} are assured to have a
reasonable number of stars and marginal contamination.

\subsection{Completeness tests}
\label{sec:completeness}

\begin{figure*} 
\center
\includegraphics[trim= 1.2cm 6.cm .5cm 5.cm, clip=true,width=1\columnwidth]{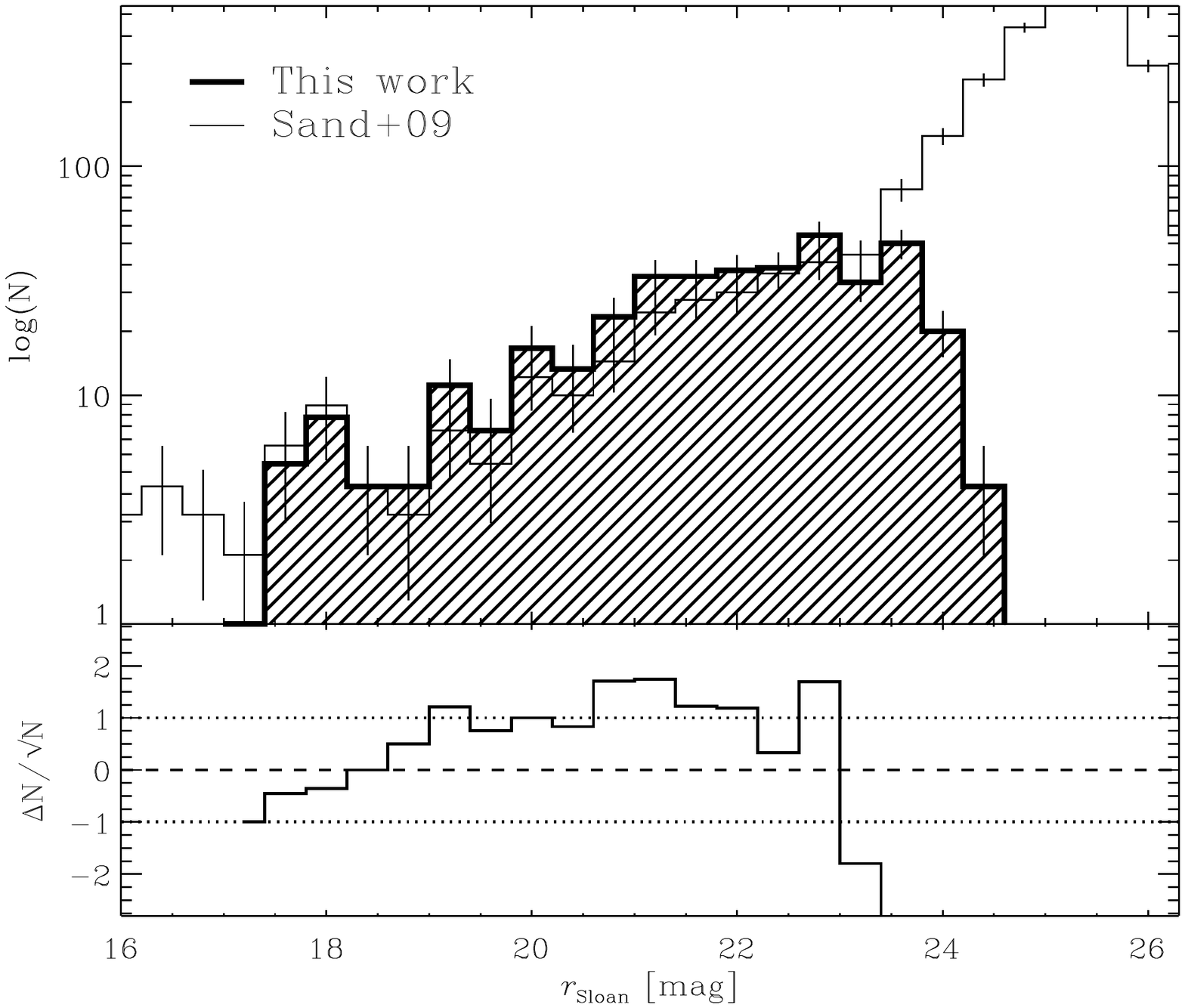}
\includegraphics[trim= 1.2cm 6.cm .5cm 5.cm, clip=true,width=1\columnwidth]{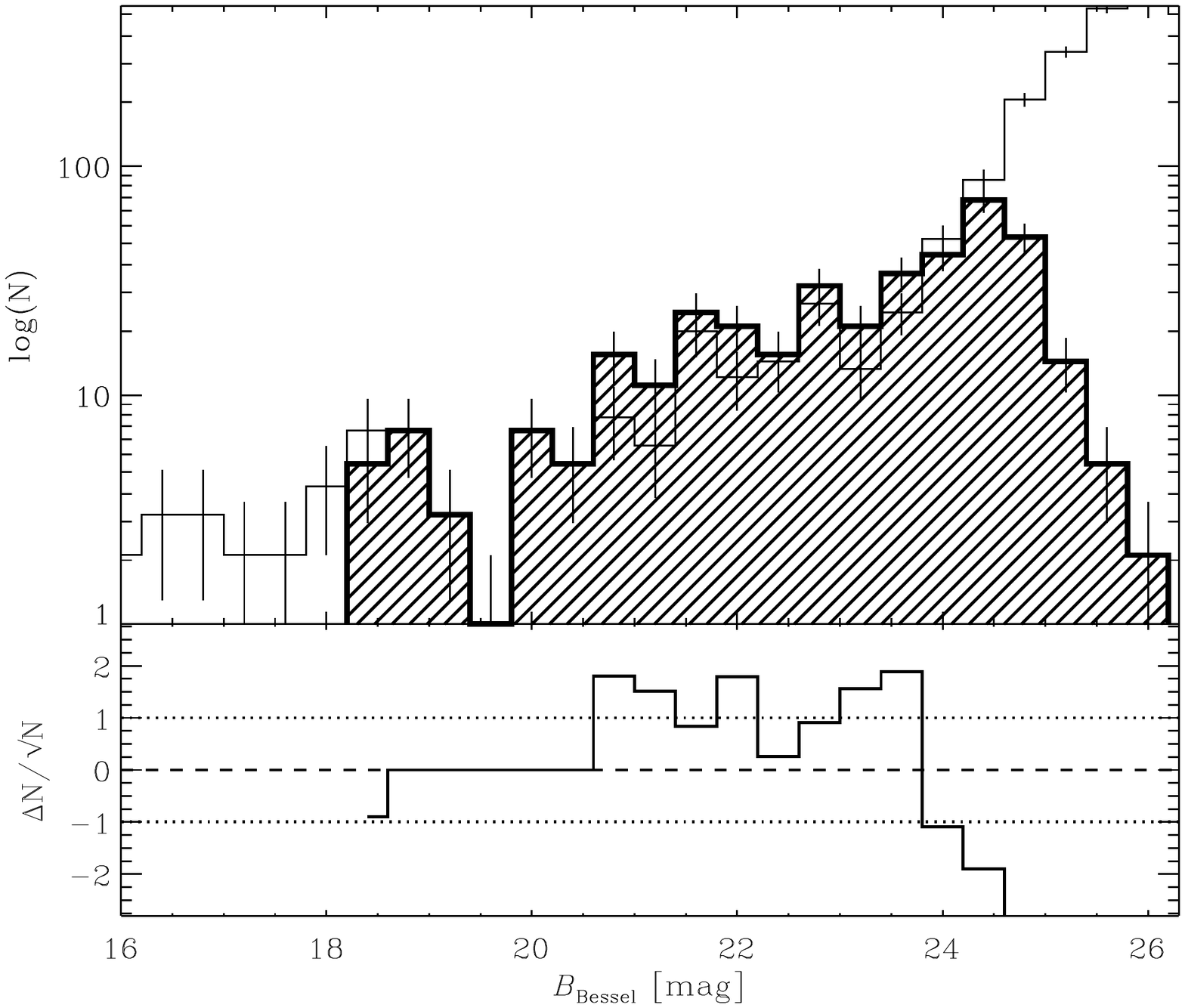}
\caption{Top panels: Luminosity functions in the \rr\ (left panel)
 and \bb\ (right panel) bands for our unselected photometric catalogue
 (thick histogram) and for \citet[][thin histogram]{sand09}. Bottom
 panels: Ratio between differential star counts for the two
 distributions and Poisson noise. The dotted lines refer to the
 $1\sigma$ level. \label{fig:compl}}
\end{figure*}

The comparison between our unselected photometric catalogues 
and those
of \citet{sand09} shows similar luminosity functions down to magnitudes
of \rr$\sim$23~mag and \bb$\sim$24~mag, respectively 
(Fig.~\ref{fig:compl}). We recall that the magnitude depth is limited by the
exposure times of the second-epoch observations, which are about five
times shorter than the first-epoch time exposures. Moreover, given the
different data reduction and photometric procedures, there are
differences between our stellar counts and those found by
\citeauthor{sand09} even for bright stars. However, such differences,
due to the differing photometric reductions, are of the order of the 
sampling uncertainty and fully accounted by the Poisson noise 
(Fig.~\ref{fig:compl}), since
$\Delta N$/$\sqrt{N}$$\lesssim$1, where $\Delta N$ is the difference in
counts between us and \citeauthor{sand09}. Fainter than the
above-mentioned magnitude limits, stellar counts differ by more than
$1\sigma$.
This implies that our photometry is nearly complete for sources brighter
than \rr$\sim$23~mag and \bb$\sim$24~mag, and, consequently, the
photometric completeness does not represent an issue in our study,
given the magnitude intervals for the analysed science objects.

\section{Membership analysis and comparison to previous works}
\label{sec:membership}

\begin{figure*} 
\center
\includegraphics[trim= 0.1cm 0.2cm 1cm 7.5cm, clip=true, height=1\columnwidth]{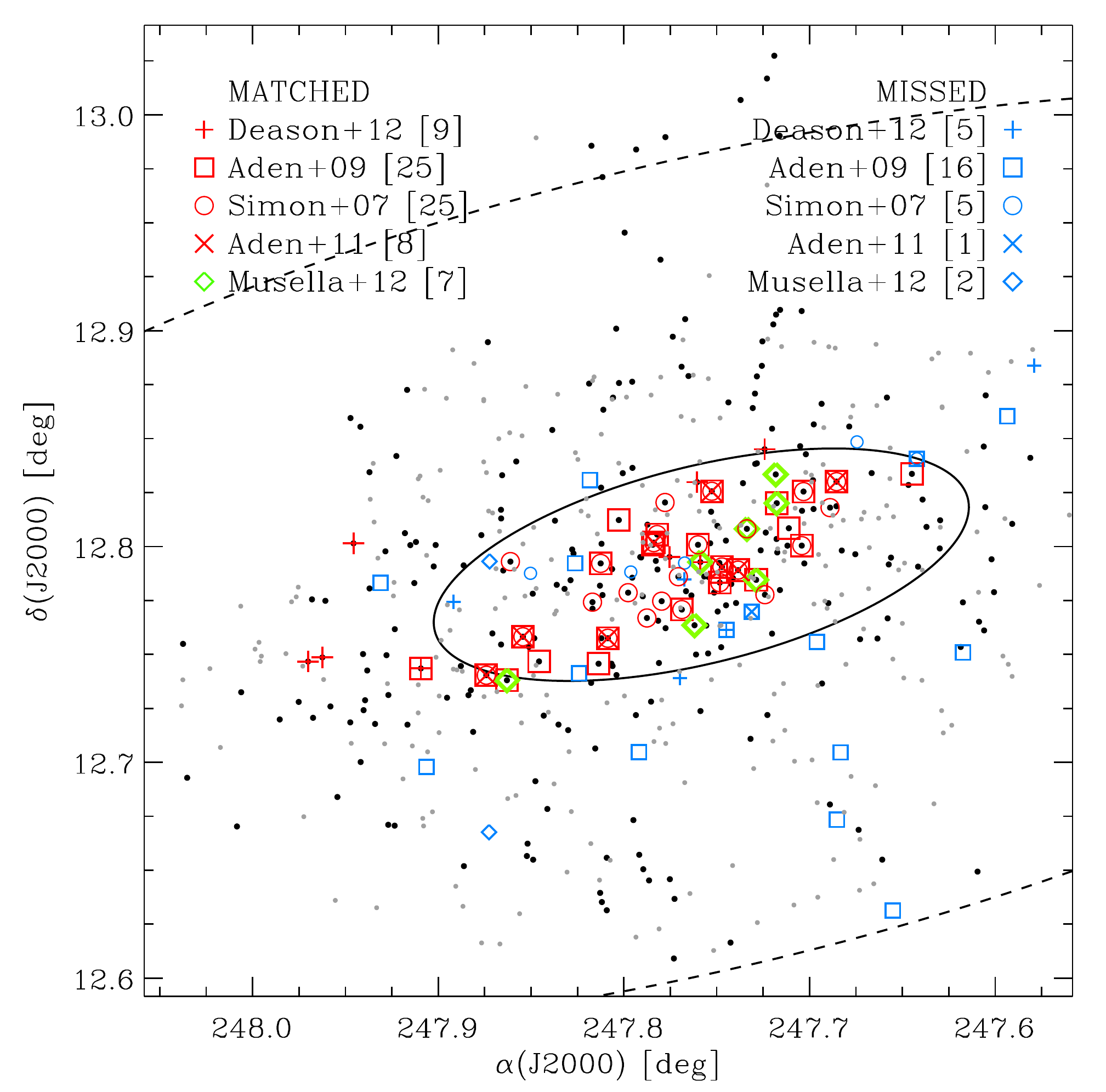}
\includegraphics[trim= 0.1cm 0.2cm 6cm 7.5cm, clip=true, height=1\columnwidth]{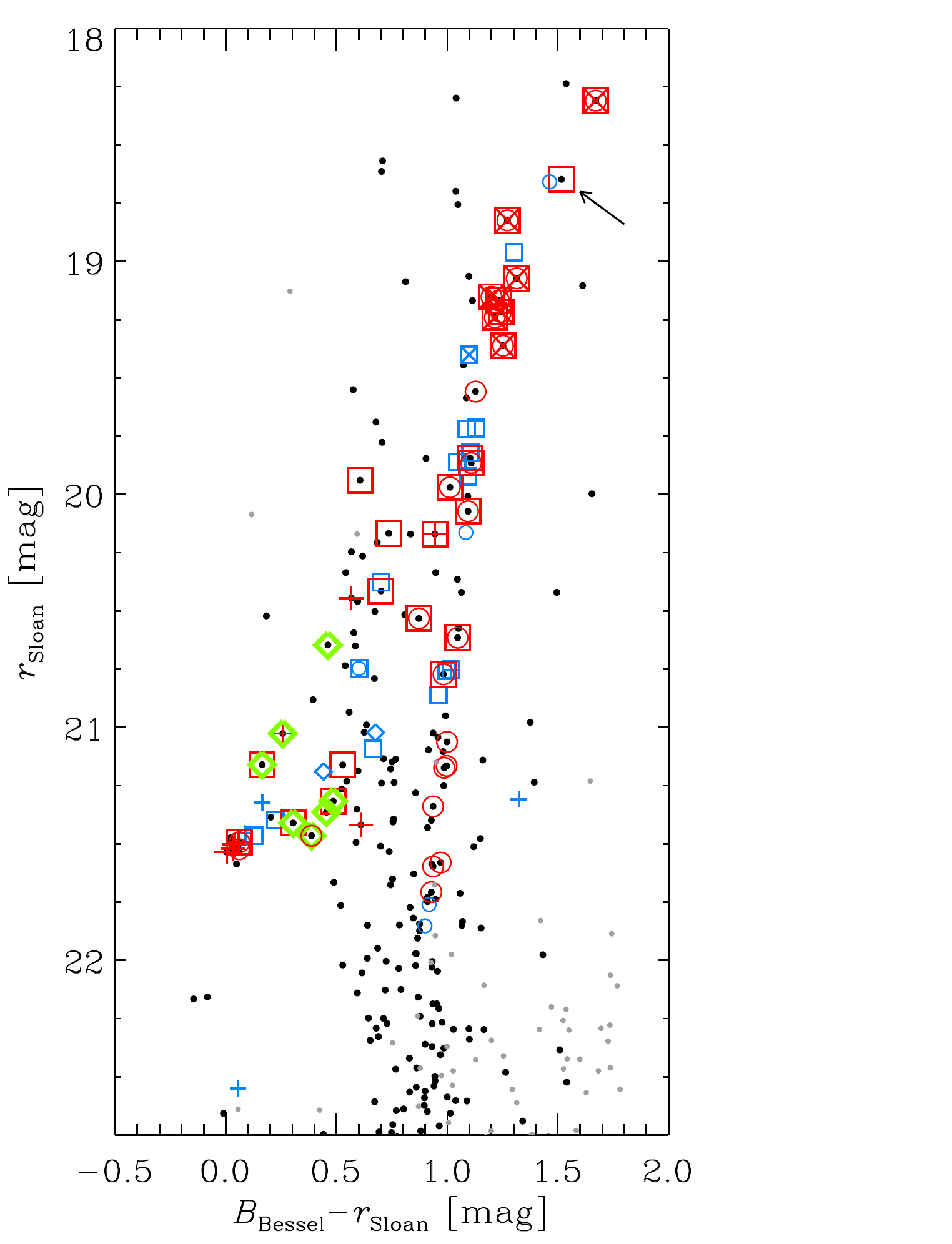}
\caption{Left: Spatial distribution of Hercules PM-selected stellar 
sources (black filled circles). Spectroscopic targets are marked as 
pluses \citep{deason12}, squares \citep{aden09spec}, circles 
\citep{simon07} and crosses \citep{aden11}. Red (blue) symbols 
represent spectroscopic targets cross-identified (non-cross-identified) in 
the PM-selected catalogue. RR Lyrae stars studied by \citet{musella12} 
are marked as green (blue) diamonds. The numbers in parentheses are 
referred to the numbers of spectroscopic targets and variables cross-
identified (non-cross-identified). Sources excluded by the colour-colour 
selection are shown as grey dots (see Fig.~\ref{fig:colcol}). Right: PM-
selected \rr\ vs. \bmr\ CMD. Symbols are the same as in the left panel. 
An arrow indicates the binary star Her-3 discovered by \citet{koch14}.
\label{fig:spatial}}
\end{figure*}

To date, attempts to identify Hercules members have been mainly
based on CMD-selection techniques, sometimes combined with
radial velocity studies. \citet{coleman07} selected Hercules candidate
members by comparing the \vv\ vs. $c_{1}$ CMD (where
$c_1=0.944[B-V]+0.330[V-\rr]$) of the galaxy central region with the
CMD of an adjacent field. \citet{aden09spec} separated the
foreground Galactic dwarf stars from
Hercules RGB giants by means of Str\"{o}mgren photometry and the
$c_{1,0}$ vs. $(b-y)_0$ diagram as a function of magnitude.
A similar approach was followed by \citet{simon07} and
\citet{deason12}, who selected galaxy candidates from the SDSS data
on the basis of the colours and proximity to Hercules.

Figure~\ref{fig:spatial} shows the spatial distribution
and the CMD of our PM-selected catalogue and the
evolved stars with spectroscopic measurements.
Red (blue) symbols mark the position of the spectroscopic targets
included (non-included) in our final catalogue. The RR Lyrae stars
identified by \citet{musella12} are also marked. 
In detail, 41 sources of \citet{aden09spec}'s list have a
counterpart in our complete photometric catalogue; however, only 25
are cross-identified in our PM-selected list.
Three stellar sources in \citet{deason12} sample are located
outside our FoV. Nine of the remaining 14 sources are
present in our PM-selected catalogue. As for the sample
of \citet{aden11}, two stars are not present in our final FoV, and
we recovered eight of the nine sources, while 25 (red
circles) of the 30 sources in \citet{simon07} are part of our
PM-selected catalogue. We note that this latter sample was also
analysed by \citet{kirby08,kirby13}.
The list of cross-identified spectroscopic sources and variables is
presented in Table~\ref{tab:spect_sources}.
In a recent study, using radial velocity measurements \citet{koch14}
confirmed the star named Her-3 in \citet{koch08,koch14} (\#41082 in
\citealt{aden09spec}, \#009 in this work) to be a binary system. We
confirm that this star (marked by an arrow in the right panel of Fig.~
\ref{fig:spatial}) is indeed present in our PM-selected catalogue.

\begin{figure} 
\center
\includegraphics[trim=0.2cm 6.5cm 1.3cm 7.5cm, clip=true,width=1\columnwidth]{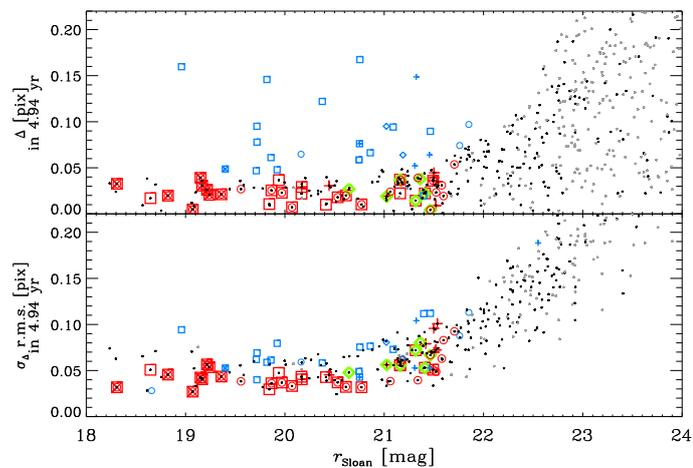}
\caption{Displacement ($\Delta$, upper panel) and displacement error
($\sigma _{\Delta}$, lower panel) as a function of \rr\ magnitude.
Symbols are the same as in Fig.~\ref{fig:spatial}. For graphical reasons,
two spectroscopic sources (non-cross-identified) are not shown in the
upper panel, since they have $\Delta$$>$0.2~pix.
\label{fig:pmerr}}
\end{figure}

The majority of the sources from \citet{aden09spec}, \citet{simon07},
and \citet{aden11} catalogues that were PM rejected are located on
Hercules RGB and HB loci (right panel of Fig.~\ref{fig:spatial}). All
these sources have a displacement larger than 0.05~pixels, where the
maximum value is allowed to be Hercules members as described in the
\S~\ref{sec:pm-analysis} (Fig.~\ref{fig:pmerr}). For
the sake of clarity, we note that seven un-matched spectroscopic
targets show a displacement larger than $0.12\pm0.01$~pixel (one
from \citealt{simon07}, four from \citealt{aden09spec}, and two from
\citealt{deason12}). Therefore, we can firmly exclude them from the
Hercules members. We also note that a 30\%\ increase of $R_{\rm
sel}$ and the adoption of $N_{\rm c}$=20, result in the inclusion of
only two stars from \citet{deason12}, four stars from
\citet{aden09spec}, one star from \citet{musella12}, and three stars 
from \citet{simon07}.

\begin{figure} 
\center
\includegraphics[trim= 0.2cm 0.3cm 1.3cm 7.5cm, clip=true,width=1\columnwidth]{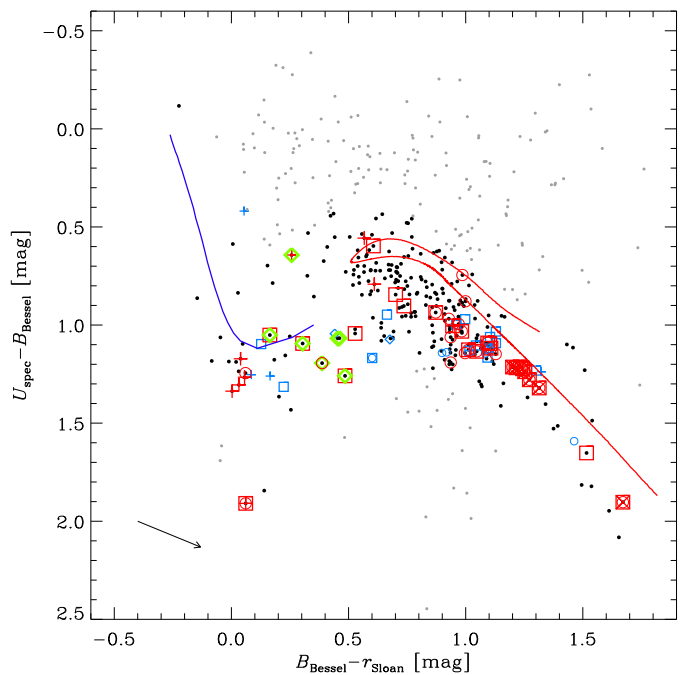}
\caption{Colour-colour diagram of Hercules PM-selected stars and
spectroscopic targets. Symbols are the same as in the
Fig.~\ref{fig:spatial}.
The isochrone with age $t$=13~Gyr and
\feh=$-2.27$ from the BaSTI database (see text) and the corresponding
ZAHB are also shown as red and blue lines, respectively. An arrow
indicates the reddening
vector.
\label{fig:colcol}}
\end{figure}

We further investigated the Hercules membership with another powerful
tool: the colour-colour diagram. In Fig.~\ref{fig:colcol}, we plot the star
distribution in the \umb\ vs. \bmr\ plane. We superimposed 
a metal-poor (\feh=$-2.27$) isochrone with age
$t$=13~Gyr and the corresponding zero-age horizontal branch
(ZAHB) from the stellar model database BaSTI
\citep{pietrinferni04,pietrinferni06,pietrinferni13}\footnote{Available at:
\tt http://albione.oa-teramo.inaf.it} as
reference. In this diagram, PM-selected stars
with spectroscopic measurements and RR Lyrae variables are located 
near the isochrone and the ZAHB, following the evolutionary phase
prescriptions also in the colour-colour diagram.
Sources located far from these regions are marked by grey dots and
are not expected to belong to the Hercules UFD. In particular, the grey
dots in the upper-right part of the two-colour diagram are mainly
background galaxies, while grey dots in the lower regions
are foreground Galactic stars \citep[e.g., see Fig.1 in][]{bono10}. In
conclusion, we are confident that stars selected according to our
PM analysis and the colour-colour diagram are likely
Hercules members.

\section{Stellar population analysis}
\label{sec:stellarpopulation}

\subsection{Metallicity distribution}

\begin{figure} 
\center
\includegraphics[trim= .3cm .5cm 1.4cm 7.5cm, clip=true, width=.99\columnwidth]{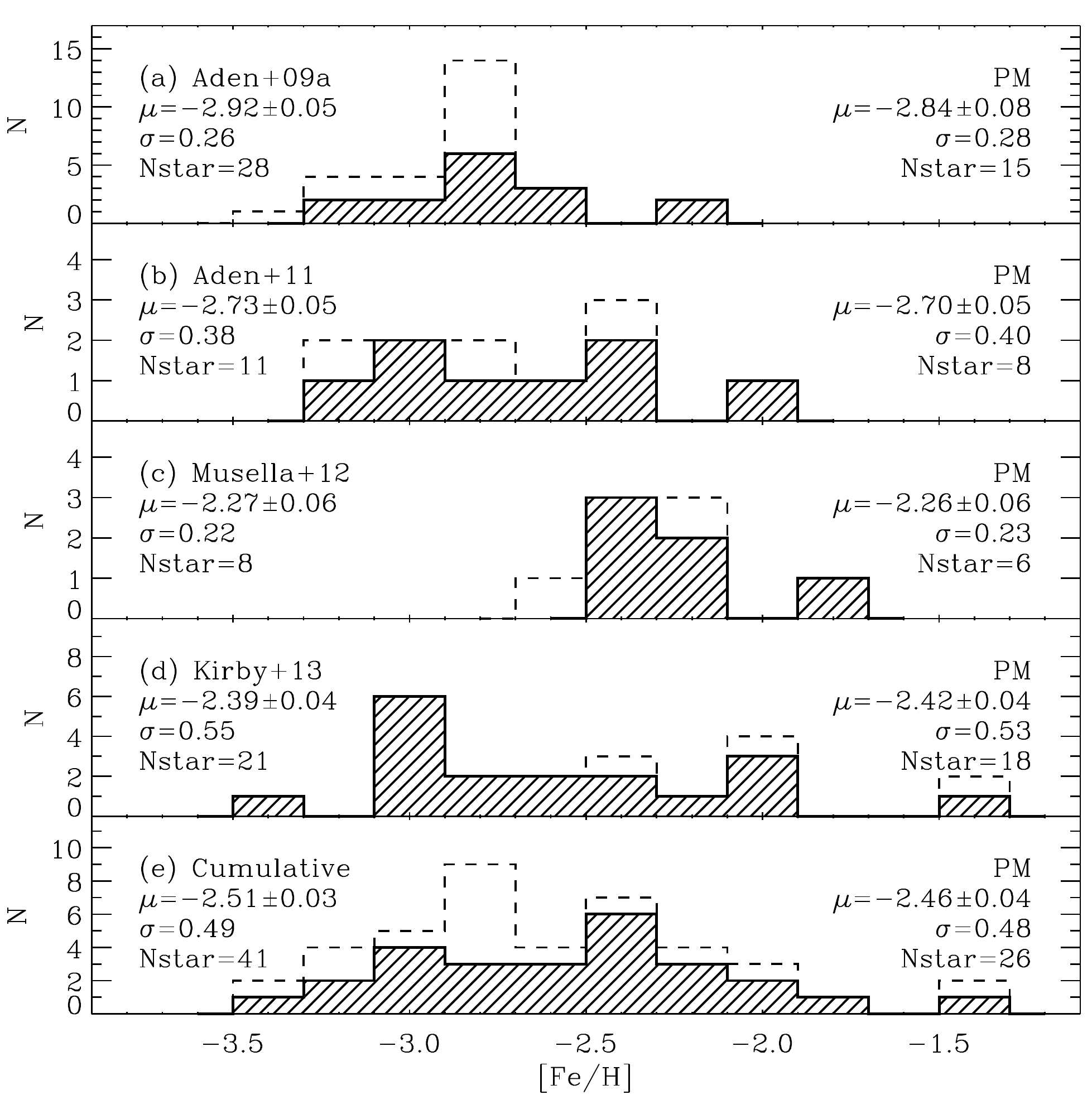}
\caption{Metallicity distribution of Hercules stars (dashed lines)
according to (a) \citet{aden09spec}, (b) \citet{aden11},
(c) \citet{musella12}, and (d) \citet{kirby13}.
The shaded histograms show the metallicity
distributions of the PM-selected stars. Weighted mean and dispersion
values are also labelled for the original (on the left) and selected (on the
right) distributions, respectively. Panel~(e): Comparison of the whole
sample of \feh\ measurements available in the literature before (dashed
histogram) and after (shaded histogram) the PM selection (see text for
more details).
\label{fig:spect}}
\end{figure}

Spectroscopic studies have shown the stars in the FoV of Hercules to
exhibit a broad metallicity distribution spanning from \feh=$-3.5$
up to $-1.5$ \citep[][see Fig.~\ref{fig:spect}]{aden11,kirby13}.
This evidence is also supported by photometric metallicity estimators,
such as Str\"{o}mgren indices \citep{aden09spec} and Fourier analysis 
of light curves of RR Lyrae \citep{musella12}.
This behavior seems quite common among dwarf galaxies (among the
UFDs, in particular) and is interpreted as the sign of a complex chemical
evolution. However, up to now, the high foreground contamination
in the Hercules field has hampered any detailed analysis. We have now
recomputed the weighted mean and the standard deviation of Hercules
\feh\ distribution using our PM-selected sample, which is only marginally
contaminated by Galactic stars and background galaxies. The metallicity
spread is confirmed, as shown in Fig.~\ref{fig:spect}, where dashed-line
histograms refer to original data, whereas shaded histograms show the
metallicity distribution of the PM-selected stars. We note that the iron
abundances, derived by \citet{aden09spec} and based on Str\"{o}mgren
photometry, were corrected by adopting the new calibration provided by
\citet{aden11}. The cumulative \feh\ distribution, obtained by merging the
three samples, is also shown in the bottom panel. From the 
PM-selected stars, a weighted mean metallicity value of 
\feh=$-2.46$$\pm$0.04 and a dispersion of $\sigma$=0.48~dex are 
obtained.

\subsection{Kinematics and structure}

We applied the same cleaning procedure to the RV distributions provided 
by \citet{aden09spec} and \citet{simon07} and obtained mean RV and
dispersion values that agree with the unselected ones. We also
analysed the cumulative RV distribution by applying a weighted mean to
the stars in common between both authors and we obtained very similar
results. Furthermore, we estimated mean systemic velocity and
dispersion using the maximum likelihood approach described in
\citet{walker06}. We derived a mean velocity of 45.02$\pm$0.97~\kms\
with a dispersion of 4.29$\pm$1.17~\kms\ using 26 stars. We used our
velocity dispersion and Eq.~11 of \citet{walker09} to estimate the galaxy
total mass for which we found the value of
$M_{\rm tot}$=3.5$\pm $2.4$\times$$10^6$~M$_\sun$, which agrees
with literature estimates \citep[e.g.][]{walker09,aden09mass}. As noted
by \citealt{aden09mass}, this value is significantly lower than the
\emph{common mass scale} found by
\citet[][$M_{\rm dyn}$=$1.4^{+0.5}_{-0.4}$$\times$$10^7$~M$_\sun$]
{strigari08}, suggesting that Hercules does not share the halo
properties seen in other dSphs.

The poor statistics of the RV distribution does not allow us to find firm
evidence for possible kinematic sub-structures in Hercules.
On the other hand, our data cover a significant
fraction of the galaxy body, and our selection procedures allowed us to
confidently identify likely members. For these reasons, we
were able to investigate the galaxy structural properties. In particular,
the left panel of Fig.~\ref{fig:spatial} shows that stars selected according
to the PM and colour-colour diagram criteria appear
strikingly distributed along the galaxy major axis. We repeated the
analysis of \citet{martin08} and confirm the large ellipticity
($\epsilon$$\simeq$0.68, $P.A.$$\simeq$$-74^\circ$) of this MW
satellite. \citet{deason12} suggested a tidal stripping process to explain
the high elongation value, a hypothesis supported by the systemic
velocity gradient in the outskirt of Hercules. We note that our PM
selection confirmed two probable tidal stripped stars \citep[][F1-09 and 
F1-10 in their Table~2]{deason12} as Hercules members. Nevertheless, 
new spectroscopic data, covering a larger area and a larger sample of 
stars, are needed to put firm constraints on the Hercules tidal disruption. 
Our study now provides a robust identification of Hercules members and 
a solid target list for further spectroscopic investigations.


\subsection{Colour-magnitude diagrams}

\begin{figure} 
\center
\includegraphics[trim= .3cm .5cm 1cm 7.5cm, clip=true, width=.99\columnwidth]{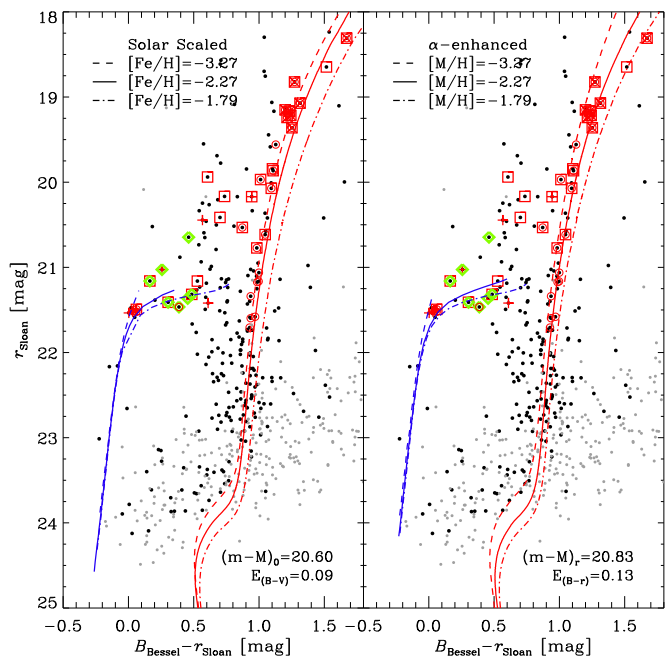}
\caption{Solar-scaled (left) and $\alpha$-enhanced (right) isochrones
overplotted to the observed \rr\ vs. \bmr\ CMD of PM and
colour-colour-selected members. Symbols are
the same as in the Fig.~\ref{fig:spatial}.
\label{fig:isoc}}
\end{figure}

Hercules appears as a very sparsely and poorly populated system (see
for instance Fig.~\ref{fig:spatial}). In spite of this, the RGB is
well-defined, and the HB appears characterised by ($i$) a group of
moderately blue stars easily recognizable at \bmr$\sim$0~mag; ($ii$) a
handful of stars that are candidate to be extreme blue HB stars; and ($iii$) 
a number of stars redder than the RR Lyrae variables and nearly at same
magnitude level. The colour-extension of Hercules HB was
interpreted by \citet{belokurov07} early as a signature of possible multiple
stellar populations in the galaxy, since a metal-rich population, possibly
1-3~Gyr younger than the bulk of stars might be responsible for the red
HB stars. Such a component was not ruled out by \citet{brown12} on the
basis of \textit{HST} photometry. However, we recall that the
morphology of the HB can be driven by a complex interplay of different 
effects \citep[e.g., see][]{lee94,buonanno97,pasquato13}.

In Fig.~\ref{fig:isoc}, we compare Hercules PM- and
colour-colour-selected CMD with theoretical prescriptions. We
over plotted isochrones and ZAHBs from the BaSTI
database for metallicities compatible with the spectroscopic estimates
(i.e. \feh=$-3.27$, $-2.27$, $-1.79$). For each metallicity, two sets of
isochrones were considered: solar-scaled and
$\alpha$-enhanced ($[\alpha/{\rm Fe}]$=+0.4)\footnote{The
evolutionary models were constructed assuming a helium-to-metal
enrichment ratio $\Delta Y$/$\Delta Z$=1.4 and a fixed mass-loss rate
described by the mass-loss Reimers parameter $\eta_{\rm R}$=0.4
\citep{pietrinferni04}.}.
An age of 13~Gyr, the distance modulus $(m-M)_0$=20.60~mag and
the reddening value $E(B-V)$=0.09~mag were adopted
\citep{musella12}. The extinction coefficients are from
\citet{mccall04}\footnote{$R_U$=4.80, $R_B$=4.07, $R_r$=2.58}.
To account for the decrease in the He-core mass caused by the use of
more updated conductive opacities, the ZAHB magnitude level was
increased by 0.05~mag \citep{cassisi07}. Agreeing with
spectroscopic measurements, we found that the RGB stars fit either
solar-scaled models or $\alpha$-enhanced chemical mixture models. In
particular, the two redder stars at the RGB-tip (\bmr$>$1.4~mag) match
well the most metal-rich isochrones.

\begin{figure*}[t] 
\center
\includegraphics[trim= 0.7cm 6cm 7.5cm 3.2cm, clip=true,height=.35\textheight]{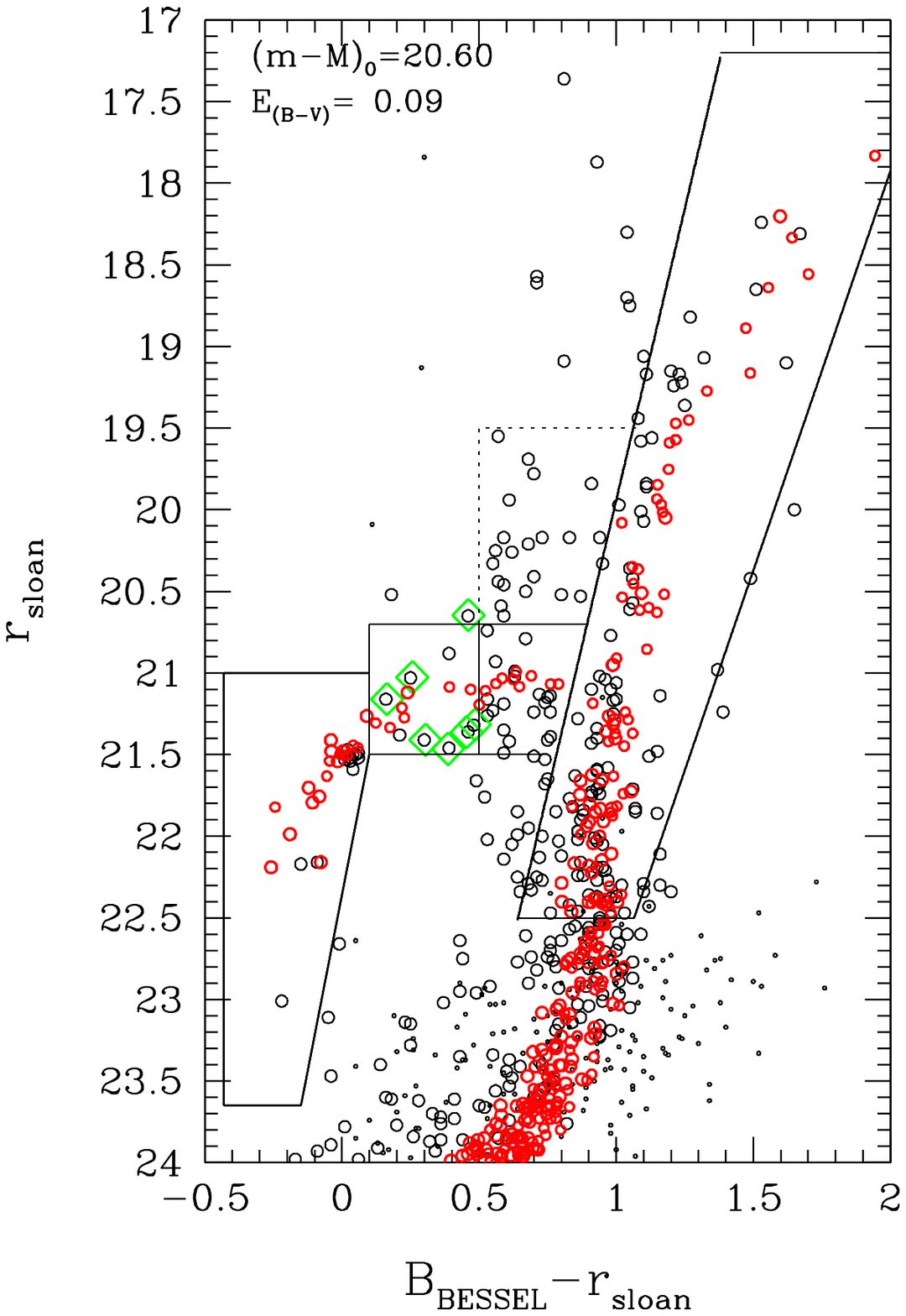}
\includegraphics[trim= 1.7cm 6cm 7.5cm 3.2cm, clip=true,height=.35\textheight]{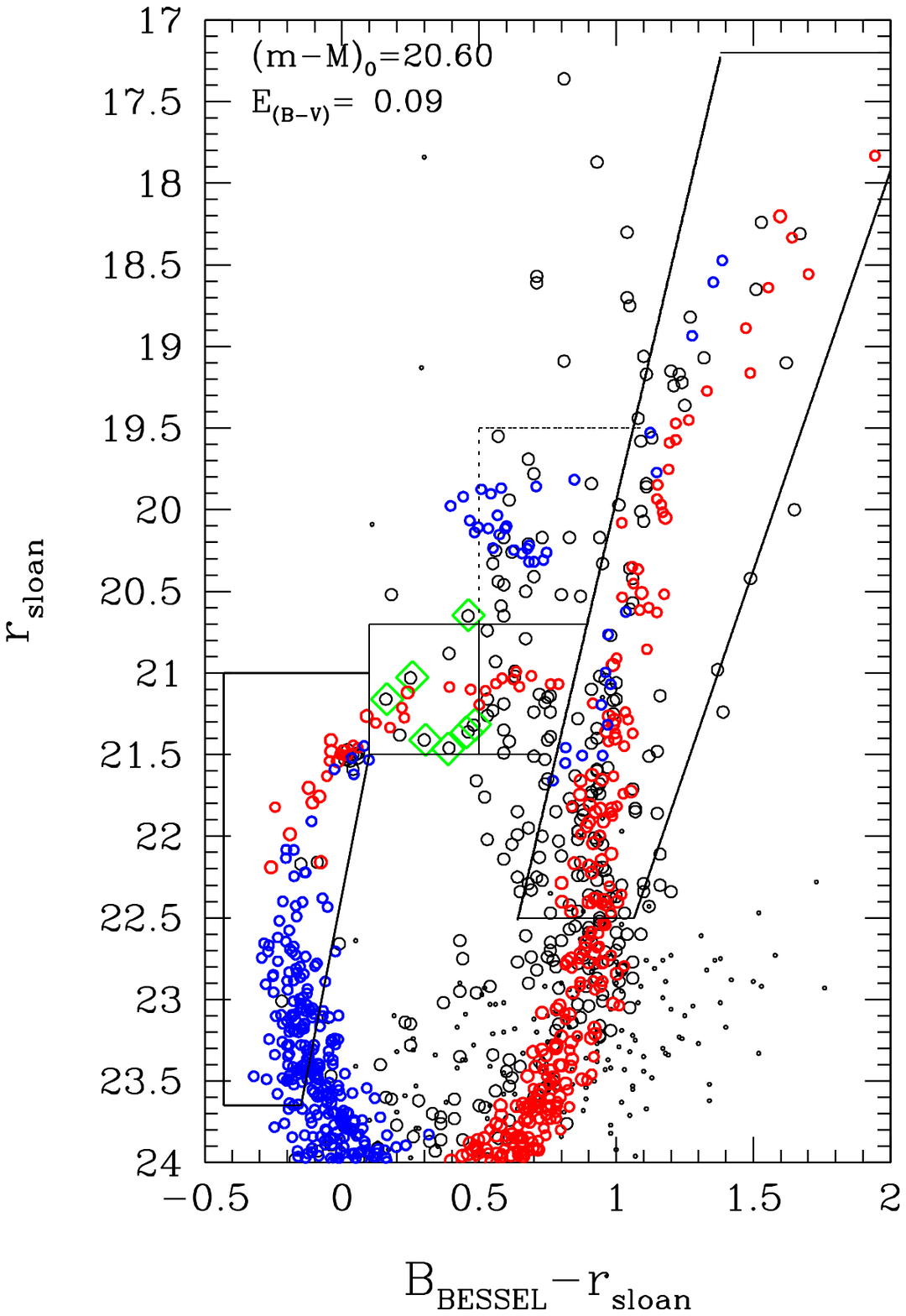}
\includegraphics[trim= 1.7cm 6cm 7.5cm 3.2cm, clip=true,height=.35\textheight]{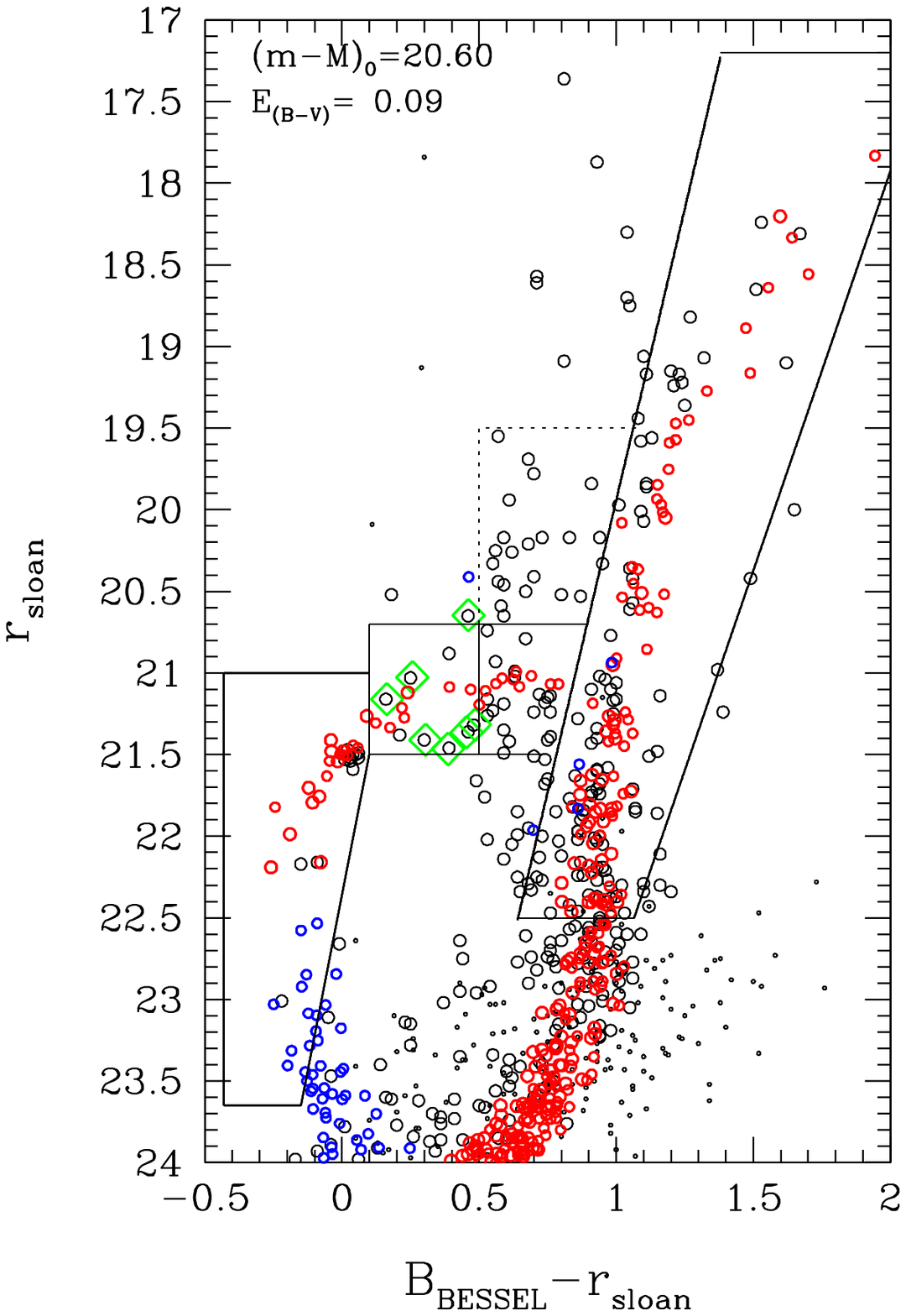}
\caption{Left: Observed CMD of likely Hercules stars (large
open black circles) compared to a synthetic CMD composed by three
stellar populations (red circles) with $t$=14, 13, and 10~Gyr and \feh=
$-3.27$, $-2.27$, and  $-1.79$, respectively. Four boxes (solid lines) are 
used to count stars on the blue, intermediate and red part of the galaxy 
HB and RGB+AGB. Observed star counts in boxes are $N$=12, 9, 20, 
and 98 (from left to right, respectively). In the dotted line box, the star 
count is 21. Observed variables are marked with green diamonds. Small 
grey circles are background galaxies and Galactic stars discarded via the
colour-colour diagram.
Middle: A stellar population with \feh=$-1.79$ and $t$=1~Gyr
(blue circles) is over plotted to obtain the same stellar count in
the dotted box.
Right: A stellar population with \feh=$-2.27$ and $t$=1.9~Gyr (blue
circles) is over plotted to obtain one star almost located in the
same position of the AC.
\label{fig:cmdbox}}
\end{figure*}

Even after cleaning for external sources, there is a number of stars
around \bmr$\sim$0.5~mag spread over one magnitude (from \rr
$\sim$20.3 to 21.3~mag). In our total FoV, we counted about 21 of such
stars (see dotted line box in Fig.\ref{fig:cmdbox}).
The feature extended above the HB level was first recognized by
\citet{aden09mass}, who also suggested it
might be populated by variable stars. However, \citet{musella12}
identified only one anomalous Cepheid (AC, star V2 in their Table~2,
\rr=20.64~mag) in that region of the CMD, which is consistent with a 
stellar evolution model with $M$$\sim$1.35~M$_\sun$ (Z=0.0001). We 
confirm that this star (the most luminous variable in Fig.~\ref{fig:isoc}) with 
a displacement $\Delta$$\sim$0~pix in 4.94~yr is a likely member
of the Hercules UFD.
The nature of ACs is still debated \citep[e.g. see][]{marconi04}. The 
most widely accepted scenarios are that they could be young 
($\leq$5~Gyr) single stars belonging to a recent star formation episode 
or they could be massive stars formed via a mass transfer in binary 
systems, but as old as the other stars in the stellar system. Therefore, 
Hercules' AC might be again an indication of the presence of an 
intermediate-age population that is as old as $\sim$2-3~Gyr, as 
speculated by \citet{musella12}.

To further investigate this issue and to study stellar
counts and distributions in the CMD in general, even in case
of a complex population scenario, the stellar population synthesis code
SPoT\footnote{\tt www.oa-teramo.inaf.it/spot}
\citep[][and references therein]{raimondo05,raimondo09} was used to
simulate a series of synthetic CMDs. We constructed a library of CMDs
for single-age, single-metallicity stellar populations spanning the
metallicity range $-3.27$$\leq$\feh$\leq$$-1.5$ and the age range 
1$\leq$$t$~[Gyr]$\leq$14. The synthetic CMDs were randomly 
populated using Monte Carlo methods with the initial mass function 
(IMF) of \citet{kroupa01} in the mass interval 0.1$\div$100~M$_\sun$. 
This procedure assures a proper study of poorly populated stellar 
systems, such as the Hercules UFD. For each star, the luminosity and 
effective temperature were derived according to evolutionary tracks by 
the BaSTI database, then magnitudes in the photometric bands of the 
LBC@LBT channels (both red and blue) were computed using 
colour-temperature relations derived from the stellar atmospheres by 
\citet[] [and references therein]{castelli03}. The LBC-filter response 
curves were taken from \citet{giallongo08} and \citet{speziali08}. We 
explored different assumptions on stellar population parameters, such
as total mass, age, metallicity, and different parameterizations of the HB
star distribution (the Reimers mass-loss parameter $\eta_{\rm R}$, its
dispersion $\sigma_{\rm R}$, etc.). Each simulation also included a
realistic photometric error.

In Fig.~\ref{fig:cmdbox}, we show the comparison between the observed
and a synthetic CMD composed by three stellar
populations with $t$=14, 13, 10~Gyr and
\feh=$-3.27$, $-2.27$, $-1.79$, respectively. The mass fraction of each
subpopulation was derived from the spectroscopic metallicity 
distribution. In the middle panel of the same figure, we also over plot a 
stellar population with \feh=$-1.79$ and $t$=1~Gyr (blue circles). From 
the simulations, we would expect a corresponding number of 
stars in the MS phase, to have 21 stars in the dotted line 
box above the red HB. It appears that \textit{HST} observations
\citep[see Fig 1 of][]{brown12} and our data, although not suitable for
studying the TO region of the galaxy due to the poor signal-to-noise,
tend to exclude such a possibility. This agrees with the general finding 
that UFD galaxies host mainly old stellar populations 
\citep{belokurov07,musella12,brown12}.
Hence, a fraction of stars in that region should belong to the MW, in spite 
of them having a displacement $\sim$0~pix in 4.94~yr (i.e. within our 
magnitude dependent PM uncertainty, see \S~\ref{sec:pm_sel}).
On the other hand, the possibility that a few stars in the aforementioned
region might be Hercules intermediate-age stars is not totally ruled out.
The right panel of Fig.~\ref{fig:cmdbox} shows the example of a
population that produces a star with photometric
properties comparable to those of the only one luminous AC known in
Hercules. This specific population has \feh=$-2.27$, $t$=1.9~Gyr, and a
total stellar mass (including white dwarfs and neutron stars see \S~
\ref{sec:hermass}) of 8$\times$$10^2$~M$_\sun$. From the figure, it is
clear that the upper part of the younger MS and the HB blue-tail of the
older population occupy the same region on the CMD, where we count
just a small number of stars. Moreover, the bulk of the younger MS stars
is fainter than \rr$\sim$23~mag, the value at which, unfortunately, our
data begins to suffer from incompleteness effects. Therefore, we cannot
exclude such a small fraction of an intermediate-age population in the
galaxy. We emphasize that the very low number of stars involved in the
analysis should deserve a more accurate statistical investigation. On the
other side, deeper data covering a larger portion of the galaxy are
needed to reach a firmer conclusion.
Interestingly, \citet{cusano13} recently identified a number of pulsating
variables brighter than the RR Lyrae stars in And~XIX, a new dwarf
satellite of the Andromeda galaxy, raising the question of whether this
galaxy was still forming stars up to 1~Gyr ago.


\begin{figure*}[t] 
\center
\includegraphics[trim= .4cm .5cm 1cm 7.5cm, clip=true,width=1.6\columnwidth]{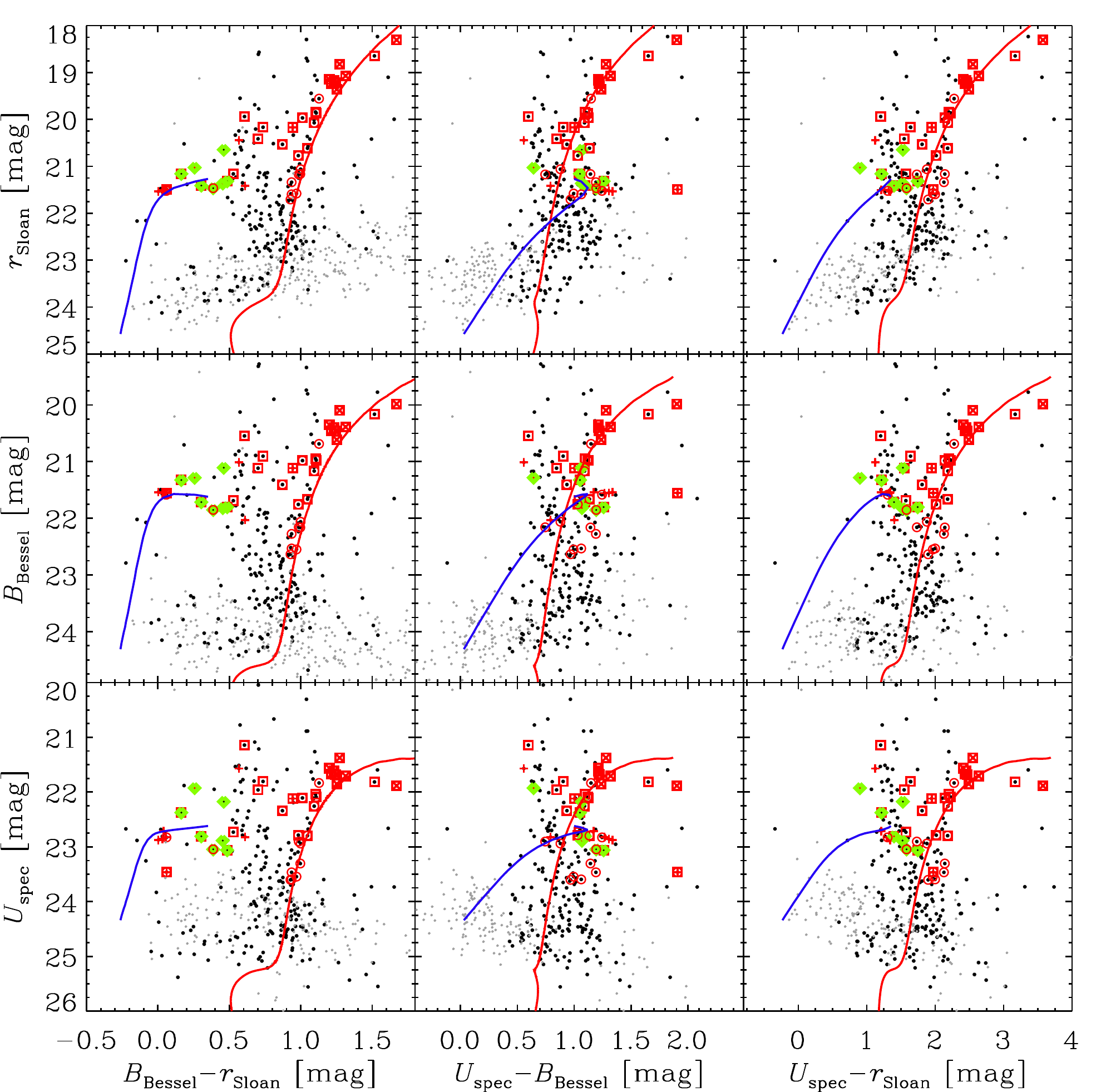}
\caption{PM-selected CMDs in different filter combinations with
superimposed a solar-scaled isochrone with $t$=13~Gyr and
\feh=$-2.27$ from the BaSTI database and the corresponding ZAHB.
The three panels on the left-hand side show, from top to bottom, the \rr, 
\bb\ and \uu\ magnitudes vs. \bmr\ colour, the middle panels show the 
magnitudes vs. \umb\ colour, and the right panels the magnitudes vs. 
\umr\ colour. Symbols are the same as in the Fig.~\ref{fig:spatial}.
\label{fig:multi_cmd}}
\end{figure*}

Figure~\ref{fig:multi_cmd} shows the PM-selected CMDs in various filter
combinations with superimposed a solar-scaled isochrone with \feh=
$-2.27$ and the corresponding ZAHB. The combination of different
bands and CMDs allow us to investigate the real membership of the
extreme blue-HB stars. We found good agreement of the blue-HB stars
with the ZAHB loci in the CMDs, thus supporting their membership to
Hercules. Interesting to note is also the HB red incursion in the panels
involving the \umb\ colour (see for a discussion \citealt{momany03}).

\begin{table*} 
\small
\center
\caption{PM-selected stars of Hercules UDF.}
\label{tab:hercstars}
\begin{tabular}{ccccccc}
\hline
\hline
ID & $\alpha$ (J2000) & $\delta$ (J2000) & \uu & \bb & \rr & $\Delta$\\
 & \tiny(deg) & \tiny(deg) & \tiny(mag) & \tiny(mag) & \tiny(mag) & \tiny(pix in 4.94 yr)\\
\hline
001 & 247.8386 &  12.8540 &  18.92$\pm$0.03 &  18.17$\pm$0.01 &  17.36$\pm$0.02 &   0.06$\pm$0.09 \\
002 & 247.8689 &  12.7897 &  18.14$\pm$0.01 &  18.14$\pm$0.01 &  17.84$\pm$0.02 &   0.02$\pm$0.06 \\
003 & 247.9269 &  12.7312 &  19.81$\pm$0.02 &  18.80$\pm$0.01 &  17.88$\pm$0.01 &   0.03$\pm$0.04 \\
004 & 247.9150 &  12.8006 &  21.59$\pm$0.03 &  19.77$\pm$0.01 &  18.24$\pm$0.01 &   0.03$\pm$0.07 \\
005 & 247.8111 &  12.8634 &  20.30$\pm$0.04 &  19.34$\pm$0.01 &  18.30$\pm$0.01 &   0.02$\pm$0.06 \\
006 & 247.7385 &  12.7891 &  21.88$\pm$0.04 &  19.98$\pm$0.04 &  18.31$\pm$0.04 &   0.03$\pm$0.03 \\
007 & 247.7318 &  12.7109 &  19.98$\pm$0.00 &  19.28$\pm$0.01 &  18.57$\pm$0.01 &   0.01$\pm$0.07 \\
008 & 247.5811 &  12.8411 &  20.04$\pm$0.03 &  19.32$\pm$0.01 &  18.61$\pm$0.01 &   0.04$\pm$0.03 \\
009 & 247.8457 &  12.7467 &  21.82$\pm$0.05 &  20.16$\pm$0.01 &  18.65$\pm$0.01 &   0.02$\pm$0.05 \\
010 & 247.7372 &  13.0069 &       \dots     &  19.74$\pm$0.01 &  18.70$\pm$0.01 &   0.03$\pm$0.07 \\
\hline
\end{tabular}
\tablefoot{This table is available entirely in machine-readable form in the online journal. }
\end{table*}

\subsection{Hercules stellar mass}
\label{sec:hermass}

We have used our PM-decontaminated sample to estimate Hercules'
stellar mass. The number of member stars above \rr$\geq$22.5~mag
(about half a magnitude brighter than our completeness limit) is quite
firm, since the contamination by Galactic stars and background galaxies
is only a few percent at this magnitude level (\S~\ref{sec:pm-analysis}).
By contrast, in the region around \bmr$\sim$0.5~mag and above the red
HB, there is still a significant field star contamination, as discussed in
\S~\ref{sec:pm_sel}. We considered stellar counts in the different boxes
shown in Fig.~\ref{fig:cmdbox} and restricted our analysis to the area
within the galaxy \rh. The observed star counts in the different boxes are
$N$=8, 6, 4, and 57, which move from blue to red in the CMD, specifically 
the red giants are $N_{RGB}$=57.

For each set of parameters (age, chemical composition, IMF, HB 
parameters, etc.), we constructed a grid of simulated CMDs by 
changing the total stellar mass to count the desired number of 
giant stars ($N_{RGB}$) in a certain number of simulations. In 
particular, the mass value changes from a few tens of solar masses up 
to $M_*$=5$\times$$10^4$~M$_\sun$. The former value is appropriate 
to produce very few red giant stars if Hercules is considered to host 
multiple generations of stars with different metal contents (and 
ages), as suggested by both photometric and spectroscopic 
observations. The lowest $N_{RGB}$ (i.e. $N_{RGB}$=0$\div$5) 
considered in the simulation is related to a generation of stars with 
\feh$\sim$$-1.79$, contributing with a mass fraction as low as a few 
percent of the total stellar content according to the spectroscopic 
measurements (see the previous section). The higher mass value 
corresponds to the mass of a single-age and single-metallicity 
population accounting for 100\%\ of observed red giants (i.e. 
$N_{RGB}$=57$\pm$5). For each set of parameters and 
each mass value, we generated 500 
simulations. Then, we selected only those simulations with a number 
of giant stars equal to the observed $N_{RGB}$. 
For each set of simulations computed with the same set of parameters 
and the appropriate number of reference stars ($N_{RGB}$), we 
computed the median of the distribution of the integrated total stellar 
mass ($M_*$), the luminosity ($L$) and the magnitudes. This procedure 
assured us to accurately recover the mass and luminosity of this sparsely 
populated faint satellite, whose luminosity can be affected by the 
number and luminosity of individual stars, which is close to the RGB-tip.
The total mass included all the evolving stars from the hydrogen-burning
limit up to the asymptotic giant branch and stars along the white
dwarf (WD) cooling sequence. In addition, stars with an initial mass higher
than $M_{up}$\footnote{$M_{up}$ is defined as the critical stellar
mass over which carbon ignition occurs in non-degenerate
conditions, marking the boundary between intermediate-mass
and massive stars. $M_{up}$ is $\sim$6.0~M$_\sun$ for the used
metallicities \citep{pietrinferni04}.}
were assumed to leave a neutron star (NS) as remnant.
To evaluate the maximum contribution of NS stars to the total mass, we
set the NS mass at its upper limit ($\sim$2~M$_\sun$, see e.g.
\citealt{Fagiolini07}).
We found that the NS mass fraction constitutes a few percent of
the galaxy mass at most .
We did not take into account the contribution of black holes, which
assumes on the mass of the remnant and the IMF.
However, for a Kroupa-like IMF in metal-poor old stellar populations it is
expected that the mass fraction of both NS and black holes does not
exceed 5-10\%\ \citep[][and references therein]{maraston05}. This
procedure resulted in more than 2000 realizations for each set of
parameters, depending on the expected numbers of stars $N_{RGB}$,
which is over more than $10^4$ total simulations.

\begin{figure}[t] 
\center
\includegraphics[trim= 1cm 6.5cm 2cm 3cm, clip=true,width=.99\columnwidth]{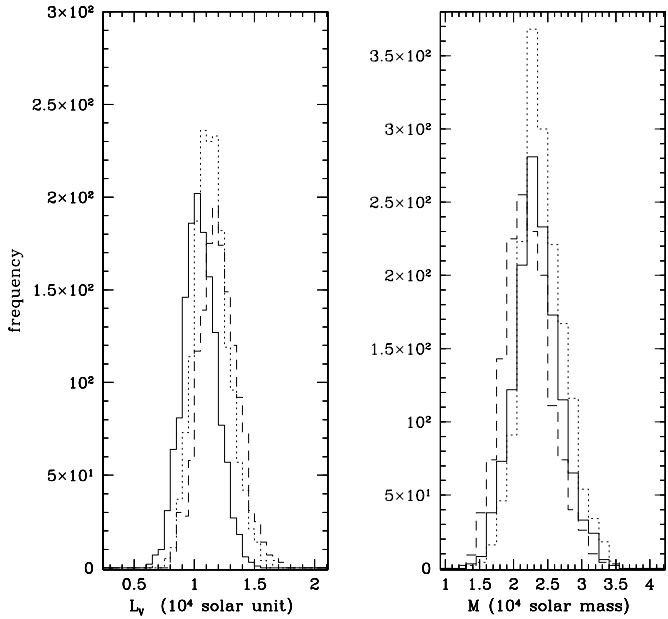}
\caption{Frequency distribution of $L_V$ and $M_*$ for simulations
containing $N_{RGB}$=57. Different lines correspond to stellar
populations with \feh=$-3.27$ and $t$=14~Gyr (solid line), \feh=$-2.27$
and $t$=13~Gyr (dotted line), and \feh=$-1.79$, $t$=10~Gyr (dashed
line).
\label{fig:stats}}
\end{figure}

In Fig.~\ref{fig:stats}, we show, the distributions of $L_V$
and $M_*$ obtained from simulations containing $N_{RGB}$=57, and
stellar populations with \feh=$-3.27$ and $t$=14~Gyr;
\feh=$-2.27$ and $t$=13~Gyr and \feh=$-1.79$,
$t$=10~Gyr, as an example.

We assumed that Hercules hosts a single-age, single-metallicity
population. In such a case, for a population with \feh=$-2.27$, 
$t$=13~Gyr, and the Kroupa's IMF (i.e. $\alpha_{\rm Kroupa}$=2.3 for 
stars with mass greater than 0.5~M$_\sun$), we found a total stellar 
mass of $M_*$=(2.5$\pm$0.2)$\times$$10^4$~M$_\sun$. As expected, 
the mass does not significantly depend on metallicity in the \feh\ range 
considered and at fixed age (e.g. 13~Gyr), being 
$M_*$=(2.3$\pm$0.2)$\times$$10^4$~M$_\sun$ for \feh=$-3.27$ and 
$M_*$=(2.6$\pm$0.2)$\times$$10^4$~M$_\sun$ for \feh=$-1.79$. At 
fixed metallicity (e.g. \feh=$-2.27$), if the age increases to 14~Gyr, the 
mass does not change, while it decreases of 
$\sim$0.3$\times$$10^4$~M$_\sun$ when $t$=10 Gyr. We also 
explored the possibility that the exponent of the Kroupa's IMF 
$\alpha_{\rm Kroupa}$ varies: for $\alpha_{\rm Kroupa}$=1.8 and 2.8, 
we found a variation on mass values of $\sim$8\%, which is 
compatible with the 
mass uncertainty due to the small number of red giants in Hercules. 
Finally, adopting a Salpeter-like IMF (i.e $\alpha_{\rm Salp}$=2.35 
down to 0.1~M$_\sun$, \citealt{salpeter55}) results in a mass value of 
$M_*$=(4.1$\pm$0.4)$\times$$10^4$~M$_\sun$.
The luminosity is much less dependent on the population parameters.
The $V$-luminosity is 
$L_V$=(1.1$\pm$0.1)$\times$$10^4$~L$_{V,\sun}$ 
and the bolometric luminosity is 
$L_{\rm bol}$=(1.4$\pm$0.1)$\times$$10^4$~L$_\sun$; changes of 
the IMF-shape or ages do not give significant variations ($\sim$10\%). 
If we rescaled the number of RGB stars to that adopted by 
\citet{martin08} and discard WD and NS stars in our simulations, we find 
results similar to \citet{martin08}.

\begin{table} 
\small
\center
\caption{List of cross-identified Hercules spectroscopic sources and variables.}
\label{tab:spect_sources}
\begin{tabular}{ccccccc}
\hline
\hline
ID & S\&G07 & A09 & A11 & D12 & K13 & M12 \\
\hline
006 &  308205 & 42241$^a$ &  42241$^a$ & \dots &  308205$^a$ & \dots \\
009 & \dots & 41082$^a$ & \dots & \dots & \dots & \dots \\
012 &  308613 & 42149$^a$ &  42149$^a$ & \dots &  308613$^a$ & \dots \\
014 &  308686 & 41743$^a$ &  41743$^a$ & \dots &  308686$^a$ & \dots \\
018 &  308457 & 42795$^a$ &  42795$^a$ & \dots &  308457$^a$ & \dots \\
020 &  309528 & 40789$^a$ &  40789$^a$ & \dots &  309528$^a$ & \dots \\
021 &  308596 & 42096$^a$ &  42096$^a$ & \dots &  308596$^a$ & \dots \\
022 &  309397 & 41460$^a$ &  41460$^a$ & \dots &  309397$^a$ & \dots \\
023 &  309487 & 40993$^a$ &  40993$^a$ & \dots &  309487$^a$ & \dots \\
026 &  308324 & \dots & \dots & \dots & \dots & \dots \\
030 & \dots & 41737$^a$ & \dots & \dots & \dots & \dots \\
032 &  308627 & 42008$^a$ & \dots & \dots &  308627$^a$ & \dots \\
033 & \dots & 43167  & \dots & \dots & \dots & \dots \\
034 &  308674 & 41912$^a$ & \dots & \dots &  308674$^a$ & \dots \\
037 &  308671 & 41758$^a$ & \dots & \dots &  308671$^a$ & \dots \\
039 & \dots & 41401  & \dots & \dots & \dots & \dots \\
041 & \dots & 40435  & \dots & F1-16  & \dots & \dots \\
049 & \dots & 41532  & \dots & \dots & \dots & \dots \\
052 & \dots & \dots & \dots & F2-01  & \dots & \dots \\
057 &  308524 & 42621  & \dots & \dots &  308524$^a$ & \dots \\
060 &  308829 & 42637$^a$ & \dots & \dots &  308829$^a$ & \dots \\
061 & \dots & \dots & \dots & \dots & \dots & V2 \\
064 &  309655 & 41423$^a$ & \dots & \dots &  309655$^a$ & \dots \\
073 & \dots & \dots & \dots & F2-07  & \dots & V1$^a$ \\
075 &  309663 & \dots & \dots & \dots &  309663$^a$ & \dots \\
083 & \dots & 42484  & \dots & \dots & \dots & V5$^a$ \\
084 & \dots & 42550  & \dots & \dots & \dots & \dots \\
085 &  3081382 & \dots & \dots & \dots &  3081382$^a$ & \dots \\
086 &  308327 & \dots & \dots & \dots &  308327$^a$ & \dots \\
097 & \dots & 40911  & \dots & \dots & \dots & V8$^a$ \\
098 &  308477 & \dots & \dots & \dots & \dots & \dots \\
100 & \dots & \dots & \dots & \dots & \dots & V6$^a$ \\
105 & \dots & 42355  & \dots & \dots & \dots & V3$^a$ \\
106 & \dots & \dots & \dots & F1-21  & \dots & \dots \\
108 &  308190 & \dots & \dots & \dots & \dots & V4$^a$ \\
111 &  308256 & 42134  & \dots & F2-17  & \dots & \dots \\
113 & \dots & \dots & \dots & F1-10  & \dots & \dots \\
114 & \dots & \dots & \dots & F2-16  & \dots & \dots \\
117 & \dots & \dots & \dots & F2-06  & \dots & \dots \\
118 &  309470 & \dots & \dots & \dots & \dots & \dots \\
120 & \dots & \dots & \dots & F1-09  & \dots & \dots \\
122 &  3081556 & \dots & \dots & \dots &  3081556$^a$ & \dots \\
125 &  308751 & \dots & \dots & \dots & \dots & \dots \\
131 &  3081473 & \dots & \dots & \dots & \dots & \dots \\
\hline
\end{tabular}
\tablefoot{Star identity from this work (Col.1),
\citet[][Col.2]{simon07};
\citet[][Col.3]{aden09spec};
\citet[][Col.4]{aden11};
\citet[][Col.5]{deason12};
\citet[][Col.6]{kirby13};
\citet[][Col.7]{musella12}.\\
$^a$ Available iron abundance value.}
\end{table}

\section{Summary and conclusions}
\label{sec:conclusions}

We have presented LBC@LBT observations in the \rr, \bb\ and 
\uu-filters
obtained in the direction of the Hercules UFD galaxy using both arms of
the LBC. Combined with LBT archive data, the \rr-filter observations
covered a time baseline of 4.94~yr (2008-2013). This allowed us to
measure the proper motion of stars in the Hercules field over an area
about six times the galaxy half-light radius.

We derived a new precise \rr-filter geometric distortion solution for the
LBC-red. This allowed us to accurately correct star positions for
geometric distortion, resulting in a final catalogue of 5385 stars in \rr\
(and \bb) and covering $\sim$0.25~deg$^2$ of the sky. We measured star
relative proper motions with a precision better than 5~mas~yr$^{-1}$
($\sim$0.10~pix in 4.94~yr) for stars down to \rr$\simeq$22~mag.
Finally, we selected 528 sources over an area of $\sim$0.12~deg$^2$,
which is a large portion of the galaxy body.

The present procedure allowed us to disentangle a significant fraction
($>$90\%\ up to \rr$\simeq$22~mag) of Milky Way stars from Hercules
members. We further improved the CMD cleaning by selecting sources,
according to their location in the \umb\ vs. \bmr\ diagram and removing
foreground stars and fainter background galaxies. We finally obtained a
sample of 357 likely Hercules stars.

We have compared our PM-selected final catalogue with photometrically
and spectroscopically selected literature samples. Our selection criteria
allowed us to clear the literature spectroscopic samples from 
non-members; however, it did not affect the weighted mean and 
dispersion of the \feh\ and RV distributions. Our PM selection provides a 
set of robustly identified Hercules members and a new target list for 
further spectroscopic observations.

The comparison of our PM and colour-colour selected CMD with
isochrones and synthetic CMDs confirmed the presence in Hercules of
an old and metal-poor stellar population with a possible spread in
metallicity consistent with that derived from spectroscopic observations
and the age range $\Delta$$t$$\sim$1--3~Gyr. 
Unfortunately, our photometric
data do not reach the turn-off point preventing more precise estimates of
age. However, we tend to exclude, although cannot totally rule out, that
the overabundance of stars in the region above the red HB is due to the
presence in Hercules of an intermediate-age population as old as 1~Gyr.

Finally, we have proven that our procedure to estimate star proper
motions, based on the LBT, can be extended and applied to stellar
systems out to large distances with high accuracy and with significant
improvement on the membership identification.

\section{Acknowledgements}

We are grateful to the anonymous referee for useful comments
that improve the manuscript.
We would like to thank the LBC SDT team and the
INAF LBT Queue run observers.
Authors thank David~Sand who provided the license
to use the first-epoch LBC data.
We are grateful to Josh~Simon, Marla~Geha and Evan~Kirby for kindly
providing spectroscopic data of Hercules stars.
Support for this work was provided by INAF$-$PRIN$/$2010
(P.I.~G.~Clementini) and INAF$-$PRIN$/$2011 (P.I.~M.~Marconi).
M.F. thanks STScI for support as a science visitor.
M.C. has been supported by FIRB 2008 (P.I. G. Imbriani).

\bibliography{manuscriptbib}

\end{document}